%% file: main.tex
\documentclass[11pt]{article}
\usepackage{./jheppub}

\usepackage[T1]{fontenc} 

\usepackage{graphicx}
\usepackage{amsfonts,amsmath,amssymb}
\usepackage{verbatim}
\usepackage{array}
\usepackage{mathrsfs}
\usepackage{mathtools}
\usepackage{yfonts}
\usepackage{dsfont}
\usepackage{bbm}
\usepackage{colonequals}
\usepackage{amscd}
\usepackage{slashed}

\usepackage{float}
\usepackage{afterpage}

\usepackage{subcaption}

\usepackage{wrapfig}

\usepackage{suffix,mathtools,cancel,bbm}

\usepackage{multirow}

\usepackage{enumerate}

	\usepackage{tikz-cd}

\input{./auxx}

\usepackage{colortbl}
\definecolor{mygray}{gray}{0.93}

\usepackage{comment}

\usepackage{amsthm}
\usepackage{caption}
\usepackage{color}
    \definecolor{darkgreen}{rgb}{0,0.5,0}
    \definecolor{darkblue}{rgb}{0,0,0.6}
    \definecolor{purple}{rgb}{0.4,.2,0.7}

\def\la{\label}
\def\nref#1{(\ref{#1})}

\title{\boldmath  Estimating global  charge violating amplitudes from wormholes }

\author[a]{Ibrahima Bah,} 
\affiliation[a]{ Department of Physics and Astronomy,
Johns Hopkins University, Baltimore, MD 21218, USA }
\author[b]{Yiming Chen,} 
\affiliation[b]{Jadwin Hall, Princeton University, Princeton, NJ 08540, USA}
\author[c]{and Juan Maldacena} 
\affiliation[c]{Institute for Advanced Study, Princeton, NJ 08540, USA}

\emailAdd{iboubah@jhu.edu}
\emailAdd{ymchen.phys@gmail.com}
\emailAdd{malda@ias.edu}

\abstract{
We consider the scattering of high energy and  ultra relativistic   spherically symmetric shells in asymptotically AdS$_D$ spacetimes.
We analyze an exclusive amplitude where a single spherically symmetric shell goes in and a single one comes out, such that the two have different global symmetry charges of the effective gravity theory.  We study a simple wormhole configuration that computes the square of the amplitude and  analyze its properties. 
 }

\usepackage{etoolbox}
\appto\appendix{\addtocontents{toc}{\protect\setcounter{tocdepth}{1}}}

\appto\listoffigures{\addtocontents{lof}{\protect\setcounter{tocdepth}{1}}}
\appto\listoftables{\addtocontents{lot}{\protect\setcounter{tocdepth}{1}}}

\begin{document} 

\setcounter{tocdepth}{2}

\maketitle
\flushbottom


\input{./sec_introduction}

\input{./sec_oneshell}


\input{./sec_solchargeV}




\input{./sec_discussion}


\subsection*{Acknowledgments}
We would like to thank  Nima Arkani-Hamed for sharing his notes with Jared Kaplan on high energy scattering of spherical shells. We also thank Aaron Hillman for initial collaboration on this subject. In addition, we thank   Stephen Shenker, Douglas Stanford, Leonard Susskind and Zhenbin Yang for discussions and comments.  J.M. is supported in part by U.S. Department of Energy grant DE-SC0009988.  I.B. is supported in part by NSF grant PHY-2112699 and in part by the Simons Collaboration on Global Categorical Symmetries.


\appendix


\input{./sec_shell}

\input{./sec_sphericalSym}

\input{./sec_JTcheck}

\input{./sec_QNM}

\input{./sec_distanceshell}
\bibliographystyle{./ytphys}
\bibliography{CVWbib}

\end{document}

%% file: auxx.tex
\newcommand{\bn}{\begin{enumerate}}
\newcommand{\en}{\end{enumerate}}

\newcommand{\ket}[1]{|{#1}\rangle}

\newcommand{\beq}{\begin{equation}}
\newcommand{\eeq}{\end{equation}}

\numberwithin{equation}{section}

\def\bea{\begin{eqnarray}}
\def\eea{\end{eqnarray}}

\DeclarePairedDelimiterX\MeijerM[3]{\lparen}{\rparen}%
{\begin{smallmatrix}#1 \\ #2\end{smallmatrix}\delimsize\vert\,#3}

\newcommand\MeijerG[8][]{%
  G^{\,#2,#3}_{#4,#5}\MeijerM[#1]{#6}{#7}{#8}}

\WithSuffix\newcommand\MeijerG*[7]{%
  G^{\,#1,#2}_{#3,#4}\MeijerM*{#5}{#6}{#7}}

\def\ket#1{|#1\rangle}

\def \beg#1{\begin{#1}} 
\def \be{\beg{equation}}
\def \bea{\beg{eqnarray}}
\def \eea{\end{eqnarray}}
\def \ee{\end{equation}}

\def \restr#1#2{{\left.\kern-\nulldelimiterspace#1\vphantom{\big|}\right|_{#2}}}



%% file: sec_introduction.tex
\section{Introduction} \label{sec:introduction}

  We expect that global symmetries should be violated in a theory of quantum gravity \cite{Hawking:1976ra,Banks:2010zn,Lavrelashvili:1987jg,Lavrelashvili:1988jj,Coleman:1988cy,Giddings:1988cx,Abbott:1989jw,Coleman:1989zu,Kallosh:1995hi,Arkani-Hamed:2006emk,Harlow:2018jwu,Harlow:2018tng,Harlow:2020bee}. It is interesting to \emph{quantify} the amount of violation for various processes, see 
\cite{Fichet:2019ugl,Chen:2020ojn,Hsin:2020mfa,Milekhin:2021lmq,Belin:2020jxr,Daus:2020vtf,Sasieta:2022ksu} for recent discussions in this direction. 
 
    
  As a simple academic problem, we can consider a two to two {\it exclusive} amplitude at very high energies, namely, energies high enough to produce an intermediate black hole state.\footnote{We thank Nima Arkani-Hamed for insisting on this question.} This is an academic question not only because it is hard to collide particles at such high energies but also because we expect many particles in the final state, rather than just two particles. In other words, the typical process consists of two particles colliding, making a black hole and then emitting many particles as Hawking radiation. 
   Nevertheless, if we insist on producing just two particles, we expect that the amplitude will be very small. We can make a simple qualitative estimate as follows. The state with two particles is very special compared to the $e^{S(E)}$ states that a black hole with energy $E$ can have. Therefore a simple estimate for the probability is 
   \be 
   |{\cal A}|^2 \sim e^{-S(E)} \la{ExpSu} 
   \ee 
  
   In this paper we consider an even simpler version of this problem where we send in a spherical shell we call $A$ and we get out a spherical shell we call $B$. This can be viewed as the s-wave sector of the above problem. We imagine that $A$ and $B$ have different charges under a global symmetry, which could be discrete or continuous. 
   Our main result is to  identify a wormhole geometry that indeed gives us the estimate \nref{ExpSu}. 
   Since we assumed that $A$ and $B$ carry different global charges, there is no semiclassical geometry that can give us the amplitude itself. However, there is a semiclassical geometry that can give us an estimate for the square of the amplitude. 
   
\begin{figure}[h]
    \begin{center}
   \includegraphics[scale=.3]{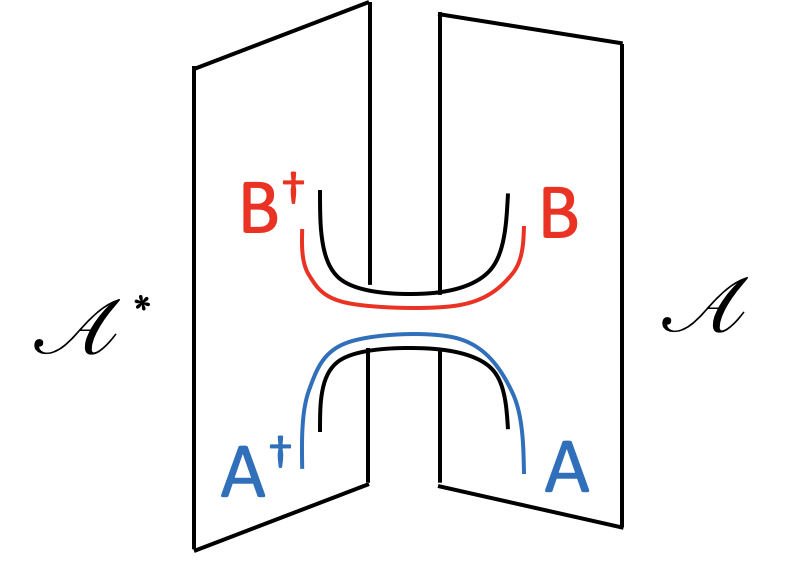}
    \end{center}
    \caption{ Sketch of the topology of the wormhole we will consider. The two sides are setting up the computation of ${\cal A}$ and ${\cal A}^*$ respectively. }
    \label{WHSketch}
\end{figure}

   This type of question is essentially the same as the one discussed in \cite{Penington:2019kki} (see also  \cite{Saad:2018bqo,Stanford:2020wkf}). Namely, their focus was the computation of an overlap between two different black hole microstates. They observed that there is a simple wormhole geometry computing the 	{\it square} of the overlap.  
    For the same reasons, the same topology appears in our problem, namely we have a wormhole connecting the regions of spacetime that set up the boundary conditions for   the computation of ${\cal A}$ and ${\cal A}^*$, see figure \ref{WHSketch}.  
    
    The main point of this paper is to discuss in more detail the geometry for  initial and final boundary conditions containing highly relativistic objects which are appropriate for a scattering situation.  The geometry contains some exotic features such as negative Euclidean time evolution which would seem to make it ill-defined according to the criteria proposed in \cite{Kontsevich:2021dmb,Witten:2021nzp}. We will argue that after including a suitable period of Lorentzian evolution the geometry appears to be sensible and trustworthy. 
   
   \subsection{The set up in more detail } 
   
   We will consider a general AdS$_{d+1}$ geometry and add operators at the boundary to produce incoming or outgoing shells
 \be \la{timesep}
 {\cal A }_{t, E, \Delta E } = \int dt' e^{ i t' E} e^{ - (\Delta E)^2 (t-t')^2 } 
 \langle 0| B(t) A(0)|0  \rangle .
 \ee 
Here an operator $A$ acts on the vacuum to produce a spherical shell with energy roughly $E$.  After a time $t$   a shell of type $B$ comes out leaving behind just the vacuum. We take $\Delta E$ sufficiently small ${ 1 \over \beta(E) } \gg \Delta E \gg 1/t $, which of course means that we consider time separations $t \gg \beta(E)$, where $\beta(E)$ is the inverse temperature of a black hole with energy $E$. 

As we mentioned above, we will be interested in a geometry as in figure \ref{WHSketch} that computes $ |{\cal A }_{t, E, \Delta E }|^2 $, in a suitable averaged sense. The simplest way to think of such an average is to average the square of the amplitude over some small window of  energies around the energy $E$. We will comment on this point a bit more in section \ref{sec:discussion}.   The assumption that we have spherically symmetric shells implies that the solution consists of several regions, each a portion of a  Schwarzschild AdS$_{d+1}$ geometry. 
 These are
 \begin{equation} \la{SchAdS}
 ds^2 = - f_\pm dt_\pm^2 + \frac{dr^2}{f_{\pm} }+ r^2 d\Omega^2_{d-1}, \qquad f_{\pm} = 1+ r^2 -\frac{\mu_\pm}{r^{d-2}}
 \end{equation} where $\Omega_{d-1}$ is the $(d-1)$ sphere.    In the $AdS_{d+1}$ vacuum region we have $\mu_-=0$, and outside the shell we have $\mu_+ \propto E$. The shell is taken to be pressure-less and composed of many massive particles.   Our main job will be to explain how the various shells connect to each other in the full geometry, providing a more detailed version of figure \ref{WHSketch} for computing the observable \eqref{timesep}. 
 
We will be interested in the regime where  the rest mass of the shell is much less than the energy $E$, so that we are dealing with highly boosted particles. In this regime, the action of the solution is dominated by some universal entropy factors and the contributions from the propagating shells is subleading. However, it is important to understand which regions of the geometry are carved out by the shells.

The paper is organized as follows. In section \ref{sec:simple}, we discus various thin shell solutions that will serve as building blocks for constructing the wormhole. 
In section \ref{sec:solchargeV}, we discuss the actual wormhole solutions that are the point of this paper. 
Finally, in section \ref{sec:discussion}, we end with some discussion and some comments on the flat space limit.   Various details that are not essential to the main point are left to the appendices.

%% file: sec_oneshell.tex
 \section{Simple shell solutions } \label{sec:simple}
 
 \subsection{One shell solution } 
 \label{sec:oneshell}

We consider a single shell in an asymptotically AdS$_{d+1}$ spacetime. This will be the basis for other solutions we will discuss later.  
 This type of solutions was discussed in \cite{Israel:1966rt} and  more recently in \cite{Keranen:2015fqa,Bezrukov:2015ufa,Chandra:2022fwi,Sasieta:2022ksu,Balasubramanian:2022gmo}.  
 For simplicity we consider a shell which is composed of a spherically symmetric distribution of massive particles, so that the integral over the sphere has rest mass $m$. 
  
\begin{figure}[h]
    \begin{center}
   \includegraphics[scale=.2]{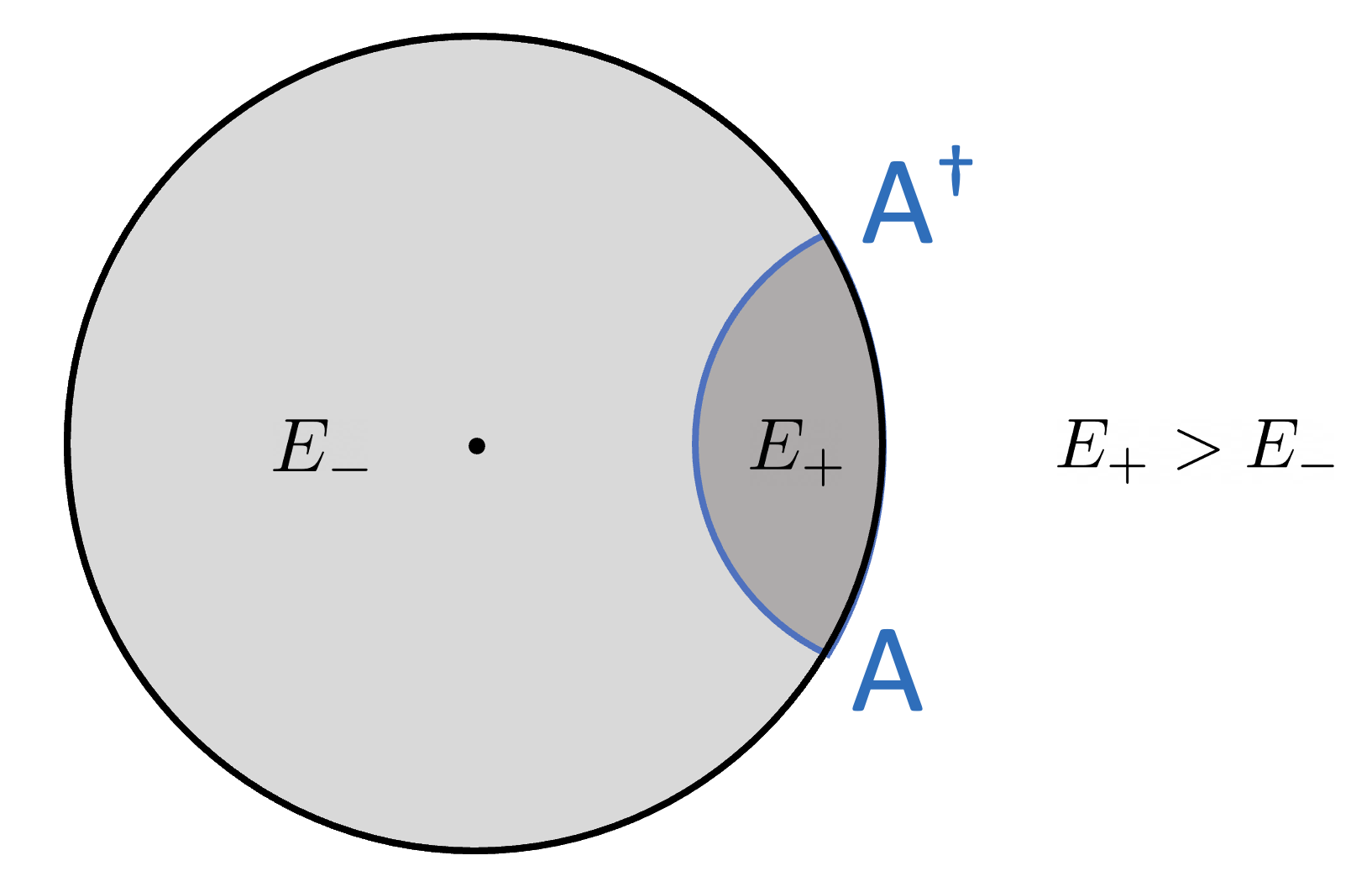}
    \end{center}
    \caption{ Single shell solution. We suppressed the $S^{d-1}$, only the Euclidean time and radial directions are indicated here. The dot at the center indicates a fixed point of the $U(1)$ isometry. }
    \label{OneShell}
\end{figure}

 The simplest solution   involves a single junction between two regions with different energies, $E_\pm$,  with $E_+> E_-$, see figure \ref{OneShell}. The metric in each region is a portion of the Schwarzschild AdS black hole metric \nref{SchAdS}, in Euclidean signature. 
 When the rest mass $m$ is small
 \be 
 \la{Lightshell} 
 E_+-E_- \gg m~,~~~~~~~~~~~~~~~~E_\pm    = { (d-1) \omega_{d-1} \mu \over 16 \pi G_N } 
 \ee 
 the solutions are relatively simple.  For this reason we will only consider this case in this paper. 
Locally, for each region of the geometry, there is an $U(1)$ isometry generated by the Killing vector $\partial_{t}$. A fixed point of this $U(1)$ isometry exists only on the portion of the geometry corresponding to the lower energy.  

 
 We can view figure \ref{OneShell} as a semiclassical estimation of   the square of the matrix element  
 \begin{align} \la{Z1}
 Z &= \int d E_- \int d E_+ \rho(E_-) \rho(E_+) |\langle E_+| A | E_- \rangle |^2  \\ &= \int d E_- \rho(E_-) \langle E_- | A^\dagger P_{E_+}  A  |E_- \rangle = Tr[  A^\dagger P_{E_+} A  P_{E_-}] \nonumber 
 \end{align}
 where the integrals are over a narrow range of energies. In the second line,  $P_{E_\pm}$ is a projector into a band of energies centered around $E_\pm$. 
   
The equations that determine the shell trajectory, in Euclidean-AdS, come from the Israel junction conditions \cite{Israel:1966rt}, which lead to, see appendix \ref{sec:shellreview}, 
\begin{equation}\la{AdSd}
\begin{aligned}
	 \dot r^2 & = f_\pm - a_\pm^2 ,\quad\quad  \dot{\tau}_\pm f_\pm =a_\pm ,\\
a_\pm & = \frac{f_- - f_+ \mp \left( \frac{\tilde{m}}{r^{d-2}}\right)^2}{2 \frac{\tilde{m} }{r^{d-2}}} =\frac{\mu_+ - \mu_- \mp \frac{\tilde{m}^2}{ r^{d-2}} }{2 \tilde{m}}, \quad \tilde{m} \equiv \frac{8\pi G_N}{(d-1) \omega_{d-1}} m,
\end{aligned}
\end{equation}
where $\omega_{d-1}$ is the volume of unit $(d-1)$-sphere. 
 
We derive the action of thin shell solutions in general dilaton gravities in appendix. \ref{sec:sphericalSym}. In terms of these coordinates, the general answer (\ref{logZ}) takes the following form 
\begin{equation}\label{actionAdSd}
	\log Z = S_- + \frac{\omega_{d-1}}{8\pi G_N} \int r^{d} (d\tau_+ - d\tau_-) + \frac{(d-2)\omega_{d-1}}{16\pi G_N} \left( \int \mu_+ d\tau_+ - \int \mu_- d\tau_- \right) - \frac{d-2}{d-1} m \int d\ell.
\end{equation} 
The first term $S_-$, being the entropy of the black hole with energy $E_-$, arises from the fixed point of the $U(1)$ isometry. The integrals are along the line describing the shell trajectory.   If we had multiple shells, we would get one such integral per shell, and we have to sum them all to get the final action. 

Let us look at the solution for the shell trajectory. 
The turning point $r_t$ of the shell  is determined by $\dot{r}=0$ in (\ref{AdSd}), leading to \begin{equation}\label{tildem}
	\tilde{m} =  r_{t}^{d-2} \left(\sqrt{f_-(r_t)} - \sqrt{f_+(r_t)} \right).
\end{equation}
For the case we are interested in, $\mu_+ > \mu_-$, and $\tilde{m}$ much smaller than the difference of the two, we can expand (\ref{tildem}) at large $\rho_t$ and get
\begin{equation}
r_t	 \approx \frac{\mu_+ - \mu_-}{2\tilde{m}} = \frac{E_+ - E_-}{m}, \la{TPsol}
\end{equation}
from which we see that the shell becomes very close to the boundary when its rest mass is small \nref{Lightshell}. 
 We also see that $m r_t$ is the energy of the shell at position $r_t$.


 In the small $m$ limit, we expect that the expression (\ref{actionAdSd})  is related to a short distance limit of a two point function. We can check it as follows. Since $r_t \gg 1$, we can use (\ref{AdSd}) and expand (\ref{actionAdSd}) in the asymptotic region
  \begin{equation}
 \begin{aligned}
 		\log Z & = S_- + \frac{\omega_{d-1}}{8\pi G_N} \int \left[ r^{d}  \left( \frac{a_+}{f_+} - \frac{a_-}{f_-}\right) + \frac{d-2}{2}\left( \frac{\mu_+ a_+}{f_+} - \frac{\mu_- a_-}{f_-}\right) -  (d-2) \tilde{m} \right] d\ell \\
 		& = S_- - 2 m \int_{r_t}^{\infty} \left( \frac{1}{r} + \mathcal{O}\left( \frac{1}{r^3}\right)   \right) dr. 
 \end{aligned}
 \end{equation}
 The integral of the $1/r$ term is divergent, but it reproduces the UV divergence from the renormalization of the operators $A$ and $A^\dagger$.
 After we subtract this UV divergence, we get
 \begin{equation}\label{smallm}
 	\log Z = S_- + 2m \log r_t + \mathcal{O}(m).
 \end{equation}
 We indeed see that in the small mass limit \nref{Lightshell} the term coming from the shell is small, so that we just get 
 \be 
 \log Z \sim S_- 
 \ee 
 which is the entropy of the lower energy solution. Note also that the second term in \nref{smallm} can be viewed as the two point function of an operator, with two insertions separated by a euclidean time  $ \tau \sim 2/r_t$, which indeed what we get by solving \nref{AdSd}. More precisely, we can think of the shell as $n$ operators, each of dimension $\Delta = m/n$,  uniformly distributed on $S^{d-1}$, in the large $n$ limit.

This same configuration can also be used to revisit a problem analyzed in  \cite{Parikh:1999mf}.\footnote{Our solution appears different from the one discussed in  \cite{Parikh:1999mf}, though the final estimate of the probability is the same \nref{ProPW}.} Namely, one starts with a black hole with energy $E_+$. The question is: what is the probability of emitting  a shell with energy $\omega = E_+ - E_- $? We imagine that the shell has  a relatively small rest mass, $\omega \gg m$. 
This computation is given by the same expression as in \nref{Z1} and figure \ref{OneShell}. The only difference is that now we need to divide by the number of initial states, $e^{S(E_+)}$,  in order to compute the typical probability in a particular initial state.  
  The final probability is 
 \be \la{ProPW}
 P \sim  e^{ S(E_-) - S(E_+) } .
 \ee 
  This example shows that we need to take care of the normalization of the initial state in order to translate the computations into physical probabilities. 
 %
  
 Note that we can view our computation as computing the final probability that a shell reaches the region far away, including all ways for the shell to get there. In particular,  the solution and its action are independent of the details of the Lorentzian solution or its approach to the near horizon region. Furthermore, in the light shell regime \nref{Lightshell} the trajectory is very far from the black hole horizon, see \nref{TPsol}. 
 We are considering a system in thermal equilibrium and the probability that we find the shell far away only depends on the total number of states that contain the shell there. It is given by a purely entropic factor. In particular, the Lorenztian problem might include possible barriers, such as angular momentum barriers, between the far away region and the horizon. Such barriers do not matter when we consider the problem in thermal equilibrium, but they are relevant out of thermal equilibrium. We comment on this point further in appendix \ref{sec:barrier}.   
   
  \subsubsection{Solution when $E_-=0$}
 
 A special case of the previous solutions arises when we set $E_-=0$, meaning that we consider an initial state with no black hole at all. 
    \begin{figure}[h]
    \begin{center}
   \includegraphics[scale=.3]{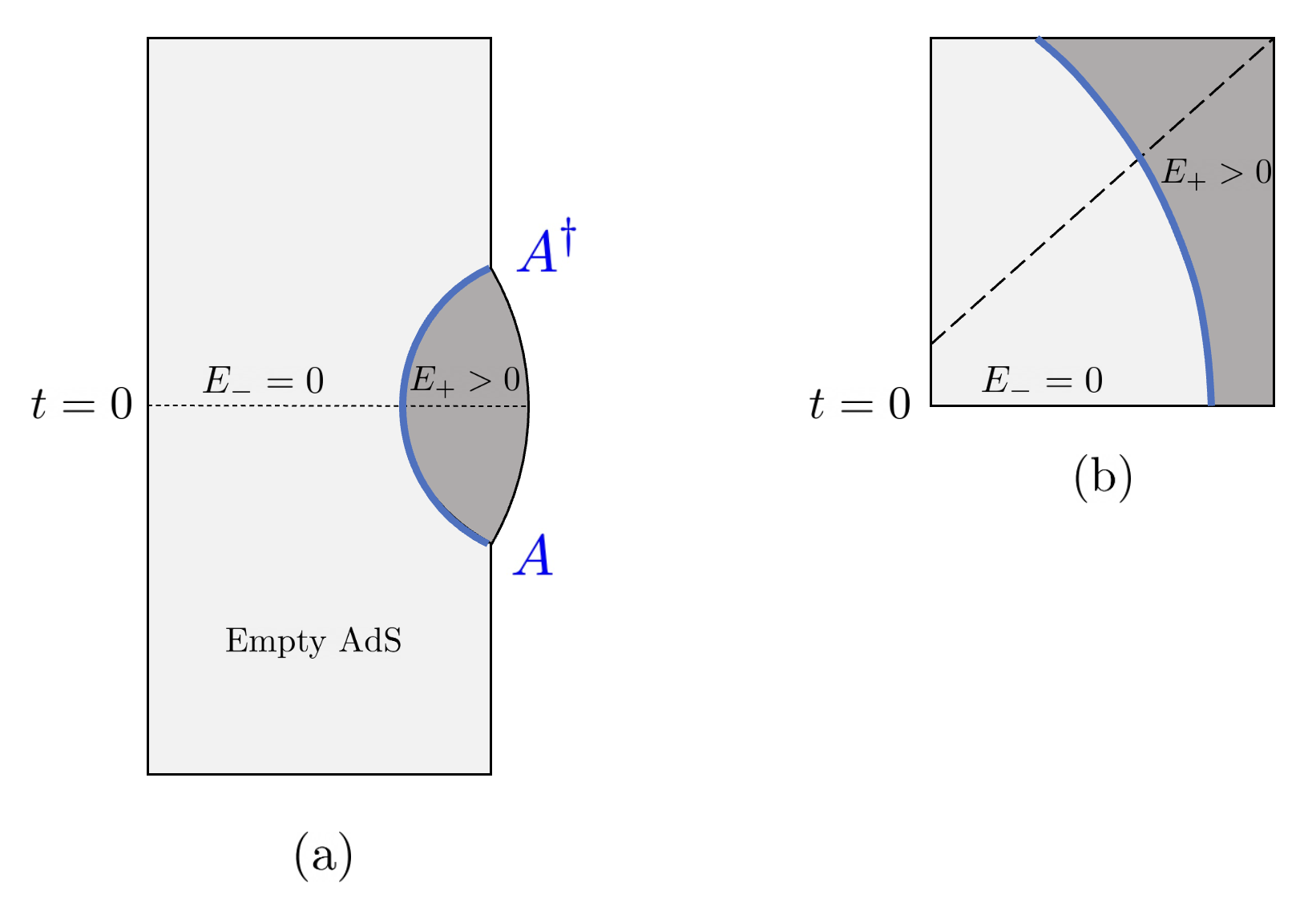}
    \end{center}
    \caption{ (a) Euclidean picture. The shell can be viewed as computing a two point function. We can view the cut along the dotted line as the state created by the operator.   (b) Picture we obtain after we evolve the state from (a) into Lorentzian time. The dashed line is the horizon.   }
    \label{VacuumOne}
\end{figure}
Then  we have $S_- = 0$ and the action (\ref{smallm}) is essentially zero in the small $m$ limit.  When $m\rightarrow 0$, we have
\begin{equation}\label{inttp}
	\int d\tau_+ = \int \frac{a_+}{f_+} d\ell = 2\int_{r_t}^{\infty}  \frac{a_+}{f_+ \sqrt{ f_+ - a_+^2 }} d\rho = \mathcal{O}\left( \frac{m}{\mu_+}\right)
\end{equation}
We should compare this shift in $\tau_+$ with $\beta(E_+)$ which is finite in the small $m$ limit. This means that   only   a very small portion of the Euclidean circle of the $E_+$ black hole is covered in this geometry. 

 Notice that we can view this configuration as the norm of a state. Namely, the state that we obtain at $t=0$  by cutting along the dotted line in figure \ref{VacuumOne}. The Euclidean evolution can be viewed as the preparation recipe for this state.    This state  can be evolved in Lorentzian time to produce a falling shell that collapses into a black hole, see figure \ref{VacuumOne} (b). The Euclidean diagram in figure \ref{VacuumOne} (a) can be viewed as computing the norm (squared) of the initial state at $t=0$. In other words, even if we do not neglect the shell contribution to the action, this shell contribution is just computing the norm of the initial state.

 
\subsection{Two shell solution}  
\la{sec:TwoShell}

Here we consider a simple two shell solution that can be interpreted as follows. We start with a black hole with energy $E$ and add an excitation $A$ that takes it to energy $E'$ and then remove it to leave a black hole with energy $E$ again. We repeat this procedure with shell $B^\dagger$. 

\subsubsection{  $E< E'$ case } 

\begin{figure}[h]
    \begin{center}
   \includegraphics[scale=.25]{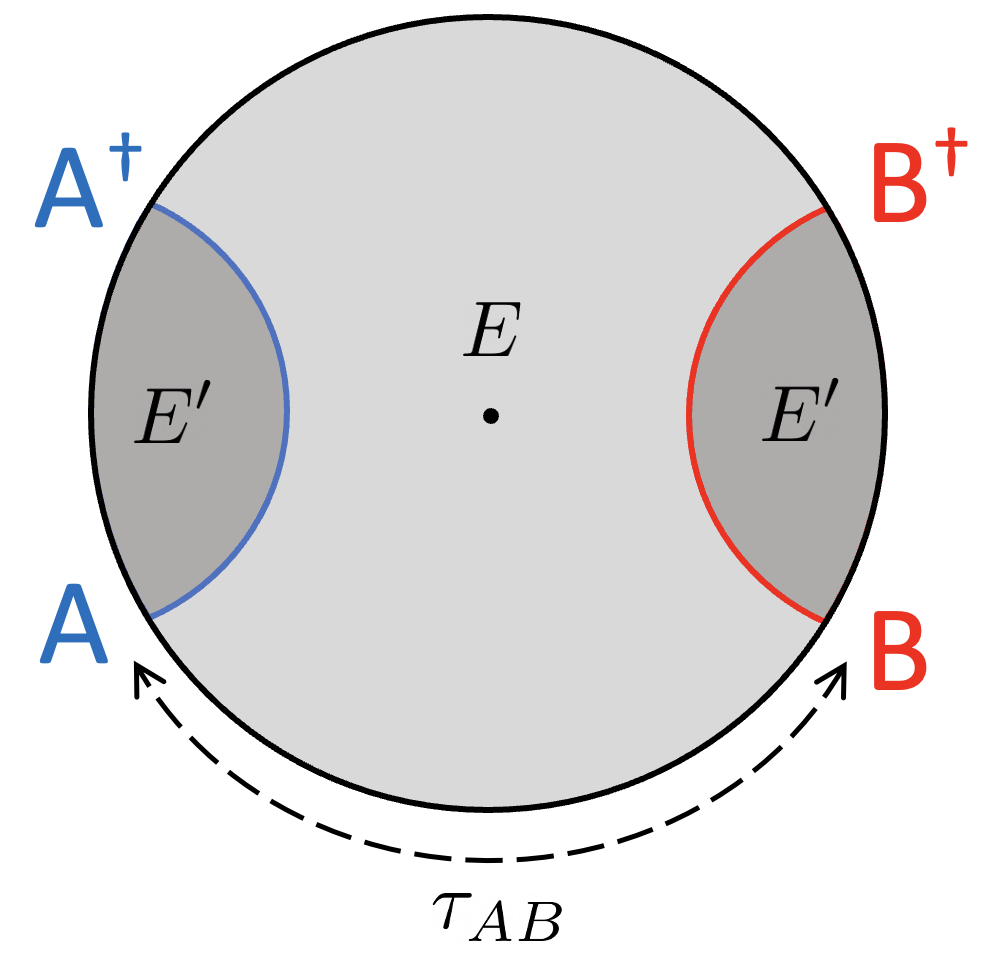}
    \end{center}
    \caption{ Two shell solution when $E< E'$. The central region, with energy $E$, contains fixed point of the isometry, but the other two regions do not contain fixed points. The operators $A$ and $B$ are separated by a positive Euclidean evolution $\tau_{AB}>0$. }
    \label{TwoEasy}
\end{figure}

The simplest version of the solution arises when $E< E'$ in which case we get the configuration in figure 
\ref{TwoEasy}. This solution contains a fixed point of the isometry only at the center of region with energy $E$, which means that the contribution is 
\be 
  e^{  S(E) } 
  \ee 
  Note the time $\tau_{AB}$ between the insertion of $A$ and $B$ is a positive Euclidean time. This time is not fixed a priori as we are fixing the energies. This time results from solving the classical equations and it represents a saddle point in the integral over times that transforms a fixed time process to a fixed energy process. 
  
\subsubsection{  $E > E'$ case } 
\la{sec:TwoHard}

\begin{figure}[h]
    \begin{center}
   \includegraphics[scale=.25]{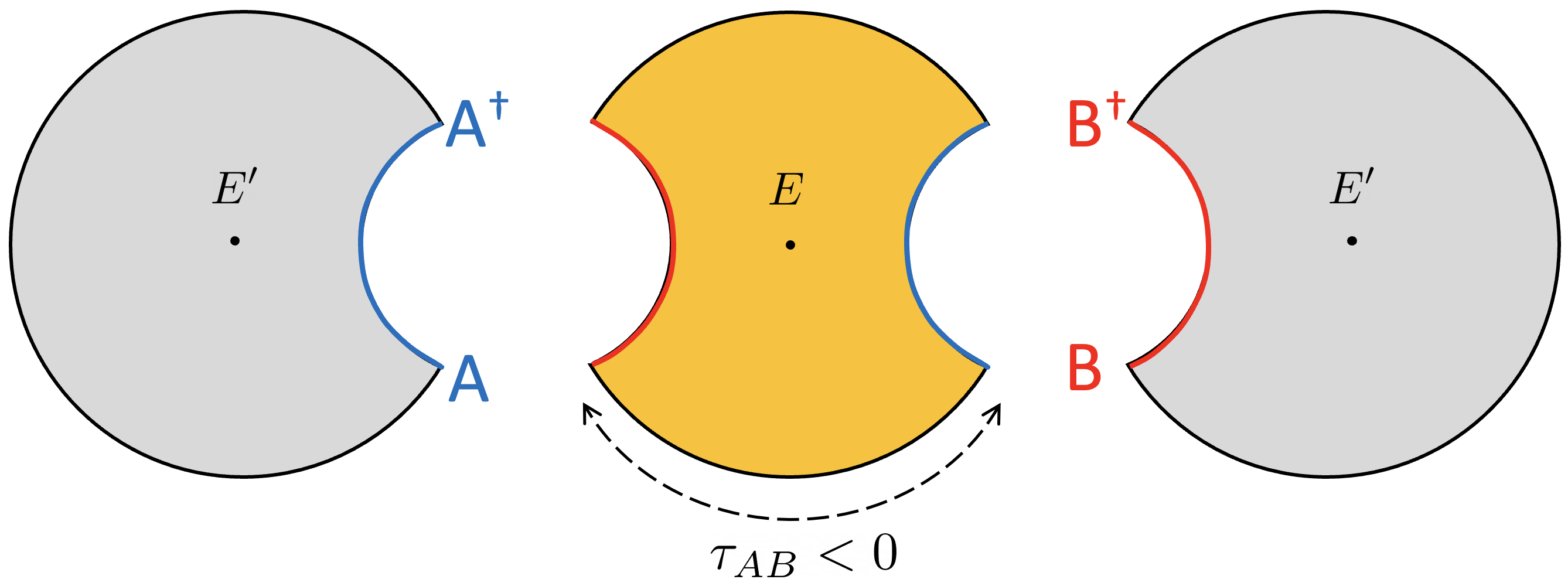}
    \end{center}
    \caption{Two shell solution for $E> E'$. The central region involves negative euclidean time evolution. Lines of the same color are identified. This means that the euclidean time separation $\tau_{AB}$ between $A$ and $B$ along the boundary of the yellow region is {\it negative}.   }
    \label{TwoHard}
\end{figure}

In this case, the solution becomes more interesting. If we follow the behavior of each shell as we lower $E'$ relative to $E$ we see that the two shells cross each other.
In fact, we would like to argue that the right solution has the form displayed in figure \ref{TwoHard}.  The interesting part is the yellow central region. We claim that in this region we have negative Euclidean time evolution, the opposite of the usual one. In other words, the Euclidean time $\tau_{AB}$ between the red and blue shell in figure \ref{TwoHard} is negative.  An explicit way to see it is to look at $\tau_{AB}(E,E')$ first in the regime $E<E'$, where it is positive, and then analytically continue it to $E>E'$. In (\ref{tauJT}) we give this function in the case of JT gravity explicitly.
Therefore, the yellow region in figure \ref{TwoHard} looks like the usual Schwarzschild solution with energy $E$, but with a Euclidean time evolution which is roughly $-\beta(E)$, the opposite to the usual one. In particular, this means that the contribution to the action that comes from the fixed point of the isometry at the center of the   yellow disk is   $ e^{ - S(E)}$, the opposite to the usual one.\footnote{This can be seen more explicitly in the discussion of appendix. \ref{sec:sphericalSym}, see eqn (\ref{EntCon}).}   For the other two disks we get a positive entropic factor so that the total contribution is 

\be \la{TwoH}
 e^{ 2 S(E') - S(E) } .
 \ee 

 We could interpret this process as one where we start from a Euclidean black hole with energy $E$ and act with an operator that lowers the energy. This is an unlikely process and we have to pay a probability price $e^{ S(E') - S(E)}$. We pay this twice since we have both the $A$ and $B$ operators. This agrees with 
 \nref{TwoH} after we divide by $e^{S(E)}$ which is the number of initial black holes. 
 Here we simply checked that the answer makes sense. 
 
 Since the configuration looks a bit exotic, it is useful to explore it in a well studied context. For that purpose we could consider this problem in JT gravity, where the exact expressions for these correlators are known \cite{Mertens:2017mtv}. We can also consider the semiclassical limit of the exact formula and see that it agrees with \nref{TwoH}, see appendix \ref{JTCheck} where we further check this for more general values of the mass.  Despite this success for the case of JT gravity,  in section \ref{Problems}  we will see that the negative Euclidean evolution poses a problem in more realistic situations.  
 
These purely entropic results for the various processes are reminiscent of similar results which involve de-Sitter tunneling events, see e.g. \cite{Blau:1986cw}.

%% file: sec_solchargeV.tex
 \section{Solutions with charge violation }  \label{sec:solchargeV}

 We now turn to the configurations that are the main point of this paper. We start first with a naive solution which has some problems but will be a good starting point for describing the proposed correct solution later.

\subsection{A naive first attempt } \label{sec:first}

Here we consider the process where we start with the vacuum, we insert an operator $A$ with energy $E$ and then we insert an operator $B$ which extracts that energy and leaves the vacuum. We insert the operator $A$ at $t=0$ and integrate over the time of operator $B$ such that the intermediate states have energy $E$.  
In other words 
\be \la{Ampl}
 {\cal A}_{A\to B} = \langle 0 | B P_E A |0 \rangle = \int dt' \,  e^{ i E t' } 
 \langle 0| B(t') A(0)|0  \rangle 
 \ee 
  where we imagine that $A$ and $B$ carry different global symmetry charges so that we do not have any bulk process going from $A$ to $B$. In other words, if we compute $\mathcal{A}_{A \to B}$ directly using semiclassical gravity, we simply get zero. Note that \nref{Ampl} is simpler than \nref{timesep} and we will discuss the importance of the difference later.  
  
  We are interested in a geometry that is a candidate for computing 
  \be \la{ProbAB}
  P_{A \to B} = |{\cal A}_{A\to B}|^2 
  \ee 

\begin{figure}[h]
    \begin{center}
   \includegraphics[scale=.35]{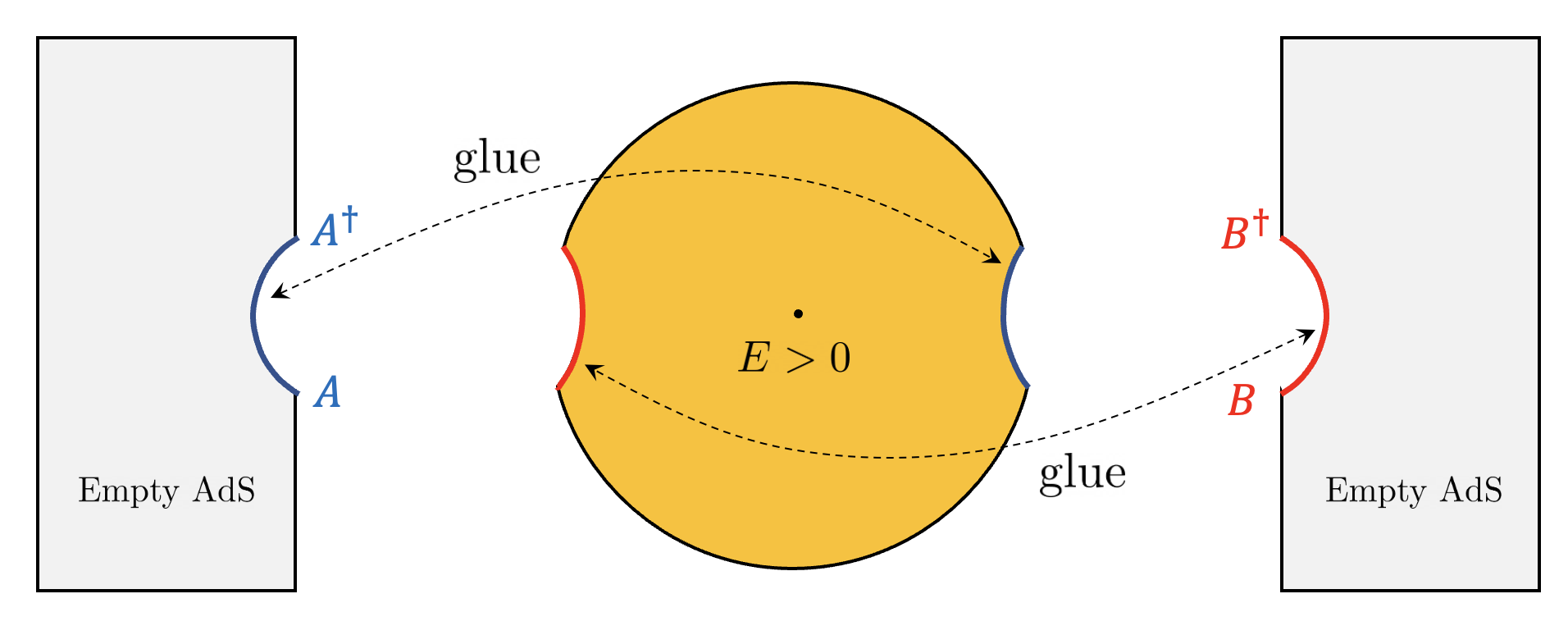}
    \end{center}
    \caption{First attempt for a solution. }
    \label{TwoShellsFirst}
\end{figure}
 
   
 The first naive proposal for geometry for computing  \nref{ProbAB} is the one specified in figure \ref{TwoShellsFirst}.   It consists of three separate segments. Two of them look like vacuum AdS and the third is a Schwarzschild AdS black hole. It is the $E' \to 0$ limit of the solution discussed  in section \ref{sec:TwoHard}. The peculiar feature of this solution, is that the boundary time evolution in the black hole region is {\it negative}. This means that the Euclidean time separation between the $A$ and $B$ operator insertions is negative. In other words, the saddle point for the time integration in \nref{Ampl} and \nref{ProbAB} is of the form 
 \be 
  t = i ( {\rm positive} ) ,
  \ee 
  where this positive constant is close to $\beta(E)/2$ where $\beta(E)$ is the inverse temperature of the black hole that we produce. This can be seen by noticing that  the black hole geometry in figure \ref{TwoShellsFirst} is given by cutting out two pieces identical to the black hole regions in figure \ref{VacuumOne}.  From equation (\ref{inttp}) we see that these latter pieces only cover a very small portion of the Euclidean circle.  Therefore almost all the Euclidean circle is left in the disk geometry, with $-|\beta(E)|/2$ on each side.   (We defined $\beta(E)$ to be positive, but we put the modulus to emphasize the negative value in our solution.)

%

  The fact that the Euclidean time is negative has an interesting consequence for the action of the solution. The  action has the opposite sign relative to the usual one for the Euclidean black hole, so that we get 
  \be 
  Z \propto e^{   - S(E) }
  \ee 
  in the regime of a light shell. This is in constrast to the usual disk contribution which contains a $+ S(E)$ in the exponent. As a special limit of section \ref{sec:TwoHard}, to understand the reason it is good to look at the derivation of \nref{EntCon} and note that we will have a length of time which is $- \beta(E)$ for the solution in question.

 \subsubsection{Problems with this solution}
  \la{Problems} 
  
   Normally we do not want to allow negative Euclidean time evolution because infinitely high energies would contribute an infinite amount. However, here we are fixing the energy, so that this problem does not obviously arise.

\begin{figure}[h]
    \begin{center}
   \includegraphics[scale=.33]{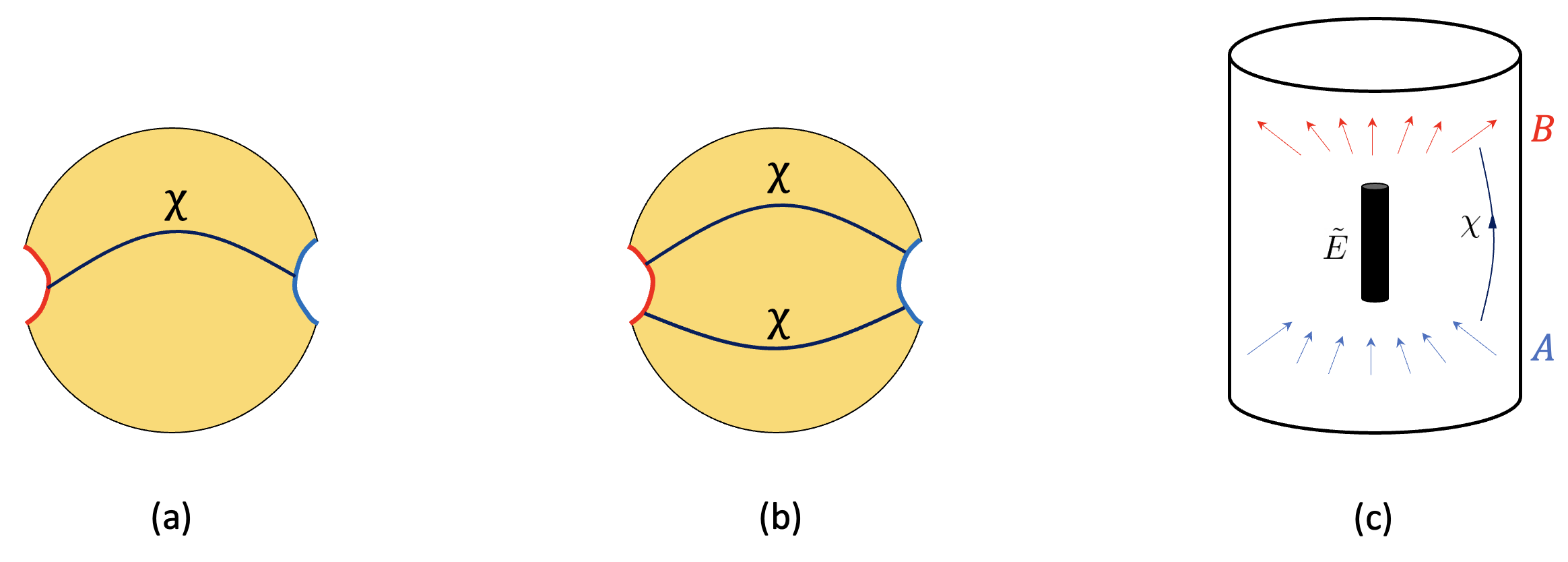}
    \end{center}
    \caption{(a) Correction due to the exchange of a single charge neutral particle $\chi$. (b) Correction due to a two particle exchange process. (c) A cartoon of the process that is captured by (b).  }
    \label{ParticleExchange}
\end{figure}

However, there is a related problem. 
  Suppose that the 
   particles created by $A$ and $B$ are coupled to another neutral particle $\chi$. This leads to the correction in figure \ref{ParticleExchange} (a), which would   give an exponentially large correction 
   \be \label{lAB}
   e^{ |\ell_{AB}| m_\chi } 
   \ee 
   where $\ell_{AB}$ is the distance between the two shells, see figure \ref{ParticleExchange}(a). 
   This makes the solution suspect and not very trustworthy. 
   
   The physical reason for this large correction can be understood as follows.   We could imagine that the shell $A$ emits a particle $\chi$ which leaves the shell with a smaller energy $\tilde E < E$. Now we can consider this shell collapsing to a black hole leaving the particle $\chi$ outside. Then we can consider a $B$ shell with energy $\tilde E $ coming out of the black hole, which then absorbs the particle $\chi$ and acquires the initial energy $E$. The diagram estimating the square of the amplitude for this process is in figure \ref{ParticleExchange} (b) while a cartoon of this process is depicted in figure \ref{ParticleExchange} (c).  The exponentially small process then happens at energy $\tilde E < E$. This smaller energy process has a suppression factor  $e^{ -S(\tilde E) }$ which is larger than the one for the case that we do not emit the particle 
   $\chi$, which was $e^{ - S(E)}$. In conclusion, the emission of the particle $\chi$ has led to an exponential enhancement of the amplitude. If the energy of the $\chi$ particle, $E_\chi$, is relatively small, $ E\gg E_\chi  $, the  net enhancement is proportional to $e^{ \beta E_\chi}$.  Of course, this result is very reasonable, since the amplitude is so strongly dependent on  the energy, we want to minimize the energy at which the charge violation happens. Unfortunately, this would take us to very small energies, where we cease to trust the solution. But this is physically saying that the most important violations (from this instanton) happen when the black hole has a size comparable to the scale of the breakdown of the effective gravity theory, which is  at  or above the Planck distance. 
      
   This same problems we discussed here would also affect the solution in section \ref{sec:TwoHard} in theories where there are interactions among the bulk particles. 
    
Another puzzling aspect of the solution is that the time separation of shell $A$ and $B$ seems to be purely Euclidean, which is in tension with the naive Lorentzian picture that the shell $B$ comes out later than the shell $A$ in Lorentzian time. We will see in section \ref{sec:improve} that the same ingredient which solves the problem of large correction also naturally solves this problem.     
    
Note that this solution does not satisfy the criterion proposed in \cite{Kontsevich:2021dmb,Witten:2021nzp} for a good complexified gravity solution. This can be seen by considering the metric close to the boundary. In the ordinary case where we have positive Euclidean evolution, the metric behaves as $ds^2 = fd\tau^2 + ...$ where $f$ is real and positive. However, negative Euclidean evolution means that we are taking $\tau\rightarrow e^{i\pi}\tau$ which gives rise to a phase $e^{2\pi i}$ in the metric. Here we are assuming that we are deforming the direction of $d\tau$ smoothly. The phase $e^{2\pi i}$ would violate the criterion proposed in \cite{Kontsevich:2021dmb,Witten:2021nzp}. 
 
  \subsection{The proposed solution } \label{sec:improve}

 In this section, we discuss a solution that seems to be under better control. To get a solution that is under control we need to modify the problem slightly.  
 
 The idea is to compute the amplitude for states where we fix the time difference between $A$ and $B$ on each side to some time $t$, but with some error, so that we can also fix the energy, also within some error consistent with the uncertainty principle. 
 In other words, we consider 
 \be \la{tAmpl}
 {\cal A }_{t, E, \Delta E } = \int dt' e^{ i t' E} e^{ - (\Delta E)^2 (t-t')^2 } 
 \langle 0| B(t) A(0)|0  \rangle .
 \ee 
 We are interested in taking $t\gg \beta(E)$, and $ 1/\beta \gg \Delta E \gg 1/t$. 
 \begin{figure}[h]
    \begin{center}
   \includegraphics[scale=.3]{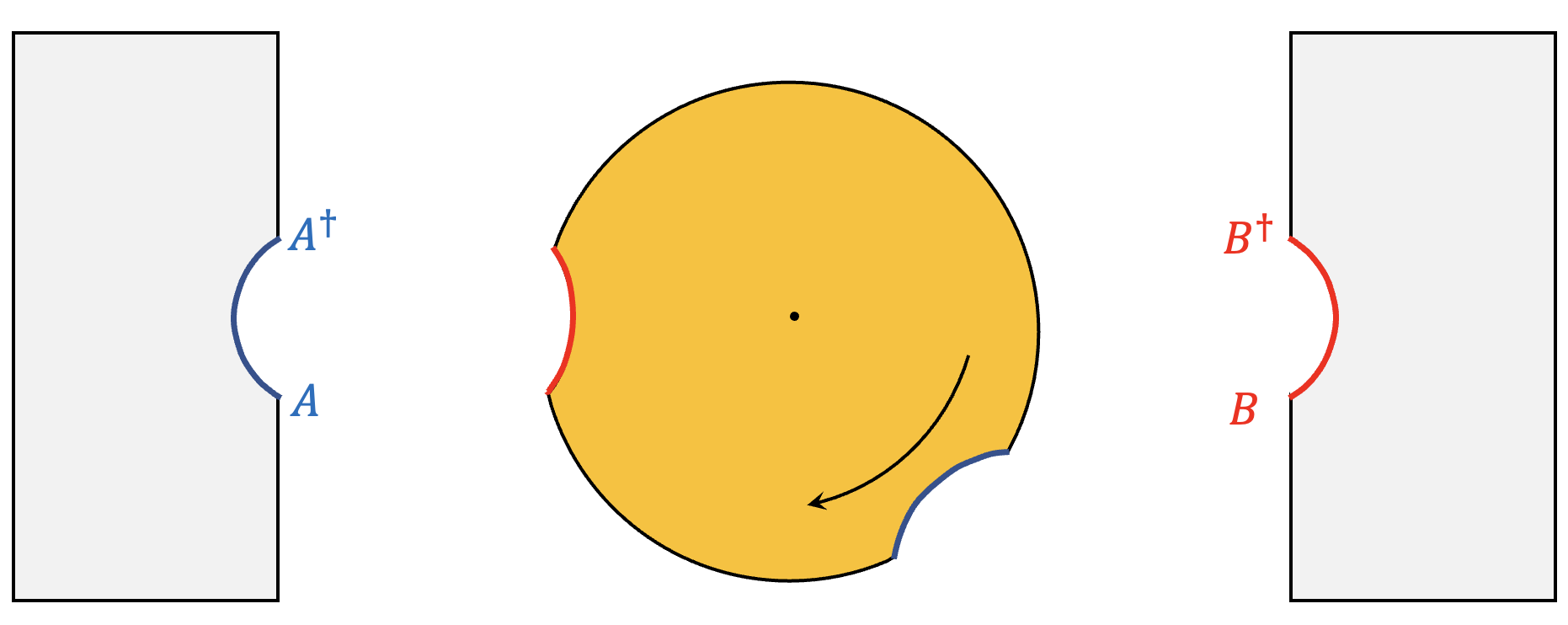}
    \end{center}
    \caption{In the shell approximation the solution has a zero mode which corresponds to a relative rotation between the two shells.  }
    \label{RelativeRotation}
\end{figure}

 It turns out that we can generate the appropriate solution from the one we already discussed above. 
 First we notice that, in the shell approximation, the solution in figure \ref{TwoShellsFirst} has a zero mode, where we rotate the position of one of the shells relative to the other, see figure \ref{RelativeRotation}. 
 This zero mode changes the time separation between $A$ and $B$. Since we want a time separation of order a Lorenzian time $t$, then we can use this zero mode, but analytically continued to Lorentzian time. This means that we are performing a large relative boost between the times associated to the shells $A$ and $B$. 
 \begin{figure}[h]
    \begin{center}
   \includegraphics[scale=.35]{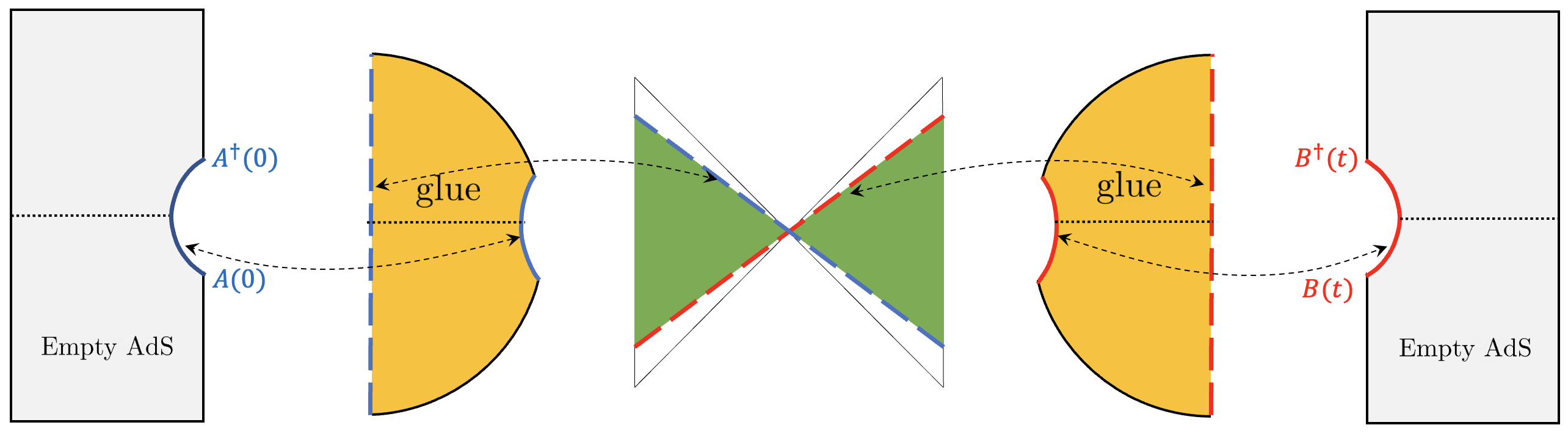}
    \end{center}
    \caption{Geometry that estimates the square of \nref{tAmpl}.    The green region is the Lorentzian ``bow tie'' geometry. The yellow region is the region with negative Euclidean time. The grey region involves positive euclidean time. Solid or dashed lines  of the same color are identified.   The dotted black line represents a cross section of the wormhole.   }
    \label{TwoShellsBetter}
\end{figure}
 This generates a large region of Lorentzian time in the solution. We call this  Lorenzian region  the ``bow tie'' geometry displayed in figure \ref{TwoShellsBetter}. This geometry consists of a bow tie section of the Schwarzschild AdS geometry, connected then to the Euclidean geometry discussed in \ref{sec:first}.   
This bow tie geometry is very similar to the geometry that appeared in \cite{Saad:2018bqo}  in connection to the ramp in the spectral form factor. The only difference is that in \cite{Saad:2018bqo} they identified the blue and red dotted lines in figure \ref{TwoShellsBetter} to form a double cone. Here instead we are connecting these to the shells.

The idea is that the extra Lorentzian evolution will essentially suppress the excitations away from the invariant states. In other words, the Euclidean evolution can be viewed as creating a certain state along the dotted lines of the bow tie geometry. There is one state along the red dashed lines and one along the blue dashed lines.   Let us call the first state $|\Psi_{AA^\dagger} \rangle_b$ and the other state $|\Psi_{B B^\dagger } \rangle_b $. The $b$ subindex emphasizes that these are states in the bulk semiclassical theory.  We are computing the overlap between these two states with the insertion of a boost in the bulk generated by the bulk boost isometry $H$ 
\be \label{amp}
 ~_b\langle \Psi_{B B^\dagger } | e^{ - i t H } | \Psi_{A A^\dagger } \rangle_b 
 \ee 
 The state that is created can be viewed as the Hartle-Hawking state plus some corrections. The Hartle-Hawking state is invariant under boosts. However, the corrections are not. In fact, 
 under large boosts we expect that the part of the overlap involving the corrections  will become very small because the states will become very different. 
 Therefore, at large times,  we expect that only the Hartle-Hawking vacuum $\ket{HH}_b$ will survive, with corrections given by the various quasinormal mode frequencies   
\be 
 _b\langle \Psi_{B B^\dagger } | e^{ - i t H } | \Psi_{A A^\dagger } \rangle_b \sim \, _b\langle \Psi_{B B^\dagger }  | HH\rangle_b ~ _b\langle HH| \Psi_{A A^\dagger } \rangle_b + \sum_n  c_n e^{ - i \omega_n t } ~, 
 \ee 
 where the quasinormal modes frequencies have negative imaginary part $\textrm{Im}(\omega_n) < 0 $ which lead to a suppression at late times. 
 
 Now, we pick a time which is large enough so that the exponential suppression due to this time evolution is larger than the exponential enhancement due to the negative Euclidean time evolution.


 The amount of enhancement coming from negative Euclidean time evolution is of order  
  $e^{\beta(E) \delta E}$, where $\delta E$ is the change of the energy of the black hole, see section \ref{Problems}.   If the real and imaginary parts of the quasinormal modes were comparable, then it would be enough to take $t$ to be somewhat larger than $\beta$ to ensure that we have a well behaved solution. 
 Unfortunately,  for AdS$_{D> 3}$  there are quasinormal modes with high angular momentum which have very small imaginary parts.\footnote{We thank Douglas Stanford for discussion on this point.}  Therefore, in order to suppress these modes we need to take times which are very large. In order to determine how large these times need to be, we need to determine the size of the imaginary part of the quasinormal mode frequencies. In the single particle approximation these can be exponentially small since we might need to tunnel through an angular momentum barrier to get to the horizon. Once we include interactions, these modes can decay by the emission of gravitational radiation that falls through the black hole horizon. These decays happen over times which are positive powers of the angular momentum $\ell$ (see recent discussion in \cite{Dodelson:2022eiz}). Now, the absolute maximum angular momentum is set by the energy of the shell. Therefore we find that the time $t$ will need to scale at most like a power of the energy, in order for the solution to be well-behaved. Of course, since the energy is large, we are talking about a very long time.

In appendix. \ref{QNM} we make some further remarks on the "bow tie" geometry. In particular, in figure \ref{TwoShellsBetter}, if we only focus on the "bow tie", the geometry has a fixed point under time evolution, which is reminiscent of the "double-cone" geometry proposed in \cite{Saad:2018bqo}. 
In appendix. \ref{QNM} we comment further on the "bow tie" geometry and its relation to quasinormal modes. 

 Another way to see that large Lorentzian time helps suppress fluctuations is to estimate the distance between the $AA^\dagger$ shell and the $BB^\dagger$ one. This distance was used in (\ref{lAB}). We discuss the calculation of this distance in appendix \ref{Distance}.  We find that a large Lorentzian time evolution gives rise to a contribution to the distance with a large positive real part,  which suppresses the particle exchange.

 \subsection{A boundary interpretation}

  Let us discuss the boundary interpretation of the bulk computation that we just discussed. This interpretation is the same as the one given in \cite{Saad:2018bqo,Stanford:2020wkf}, adapted to our setup. 
    
    On the boundary theory,  the square of the amplitude is a computation of the form 
    \be \la{AmpSq}
 |{\cal A}|^2 \sim    \int dt_R \,\nu(t_R)  \int dt_L \,\nu^* (t_L) \langle 0| B(t_R) \underbrace{ A(0) |0 \rangle \langle 0| A^\dagger   (0 )} B^\dagger(t_L)  |0 \rangle 
\ee 
where $\nu(t') = e^{ i t' E} e^{-\Delta E^2 (t -t')^2 }$. 

This computation exactly factorizes. It is interesting to think about its behavior as a function of the time $t$ which governs the rough separation between the insertion times of the two operators. It is useful to think in terms of the evolution in the doubled system, one evolving forwards in time and the other backwards in time. In other words, we can view the term with braces in \nref{AmpSq} as producing a state in a doubled system 
  \bea \la{AAdag}
A |0\rangle \langle 0| A^\dagger  &~~\longrightarrow ~~~& |\Psi_{A A^\dagger} \rangle  = A^\dagger_L A_R |0\rangle |0\rangle 
\eea
This state is completely factorized. We then  evolve it with $e^{ - i t (H_R - H_L)} $ and then overlap it with a similar ket state produce with the $B$'s. In this situation, for large $t$ we expect that the terms corresponding to the propagation of states with different energies $  |E_L - E_R|\gtrsim 1/t$ should give oscillatory phases that would average to zero. This is related to the decay of the quasinormal mode and thermalization in the boundary theory \cite{Saad:2018bqo,Stanford:2020wkf}. We also see that thermofield double like states with $E_L = E_R$ are left invariant under this evolution and will contribute to the long time average of \nref{AmpSq}.  However, just this time evolution does not generate a projection on to a TFD state since that would create entanglement and non-factorization. However, a suitable joint average over $t$, over an energy window, or over some microscopic couplings could do that. This then is interpreted as giving us the typical value of the square of the matrix element, as in \cite{Saad:2018bqo,Stanford:2020wkf}.

%% file: sec_discussion.tex
\section{Discussion} \label{sec:discussion}

 In this paper we have presented a geometry, summarized in figure \ref{TwoShellsBetter}, that computes an estimate for the square of an amplitude for the scattering of spherically symmetric shells coming and going to the AdS$_D$ boundary, 
 \be \la{final}
 |{\cal A}_{A \to B}  |^2  \sim e^{ - S(E)} 
 \ee 

 The geometry involves a wormhole and therefore all the same interpretational issues that were described previously for other cases \cite{Coleman:1988cy,Giddings:1988cx,Maldacena:2004rf} also arise in this case and we now briefly summarize them. 
 This computation is most sensibly interpreted as the result of an average over microscopic details of the quantum gravity theory \cite{Saad:2018bqo,Penington:2019kki,Stanford:2020wkf,Marolf:2020xie}. In particular, we could view the amplitude, ${\cal A}$ itself as a fluctuating quantity that is a random variable whose  root mean square is what we computed. In particular it also has a fluctuating phase. The fluctuations arise when we vary the couplings, vary the energy, or vary other detailed quantities. A concrete way to interpret the average is simply averaging the square of the amplitude over a small high energy window. By considering higher powers $|{\cal A}|^{2n}$ and looking at wormhole contributions consisting of pairs of wormholes, one can argue that the quantity $\mathcal{A}$ is a complex gaussian (in this approximation) \cite{Saad:2018bqo}. 
 
 
 It would be interesting to understand the prefactor in (\ref{final}). This would force us to understand more clearly the quantum fluctuations around the saddle point geometry. 
 
 One comment is that we do not expect to find a non-zero answer if the symmetry is gauged as opposed to global. The difference is that now there are multiple solutions that differ from the original one by the action of a symmetry operator on one of the sides. These give different configurations because they create a (flat) gauge field with a non-zero integral across the wormhole. If the symmetry is discrete there are just a discrete set of values and if it is continuous a continuous set. But the net effect is the same, these produce non-zero phases which add up to zero when there is a net charge flowing through the wormhole \cite{Coleman:1988cy,Giddings:1988cx}.

As it was discussed in \cite{Polchinski:1999ry,Susskind:1998vk,Giddings:1999qu,Penedones:2010ue},  we can extract flat space scattering amplitudes from limits of AdS correlators. This is just a limit of the correlators we have been considering. 
We might think that the limit is straightforward, and that the answer continues to be given by $e^{ - S(E)}$. 
Nevertheless, this answer is  unsatisfactory since we are not obtaining it from a solution where we are putting scattering conditions at null infinity. 
In other words, one would have liked to have a solution where we have null shells in the asymptotically flat space region. 

The difficulty with the flat space limit is also connected to the fact that we are thinking of the intermediate black hole as an object in thermal equilibrium, with radiation that goes all the way to infinity. In a scattering situation this will not be the case. 




 In flat space, it is interesting  to compute two to two amplitudes for various values of Mandelstam variables $s$ and $t$.  We can view these as resulting from higher partial waves, $j>0$, with
  Kerr black holes as intermediate states. However, the shell analysis is not as simple as for $j=0$,   since Kerr is not the unique solution outside a body with mass and angular momentum. So the configurations could be more complicated and we have not looked at it in detail.

 The problem of computing high energy scattering amplitudes using classical gravity solutions was recently discussed in  \cite{Bezrukov:2015ufa}, who also considered the scattering of shells. One difference is that they had the same shell going in and out (so no charge violation) and they used it to compute the amplitude itself. Their shells were moving into the complexified black hole geometry  and ``scattering'' off its features.   In particular, the shell trajectories are {\it not} the same as the ones we considered and there was no wormhole.   
 
 Note that we estimated the violation through this calculable process which is exponentially small in the regime where we can estimate it.  However, we expect that a more realistic theory would lead to stronger violation via Planck suppressed operators that violate the symmetry. 

  We concentrated on this particular process, just to understand the rules for how to compute it since it involves a fairly simple setup.  However, this is {\it not} the most interesting question regarding charge violation. For practical purposes what one wants to do is to put bounds on the coefficients of local operators that violate the global charge. For example, one is really interested in the dimension six operators mediating proton decay in the Standard Model.  Perhaps bounds might be obtained by some kind of dispersion relation argument which could relate the high energy behavior described here to the low energy one, but we do not see  a way to do it.

%% file: sec_shell.tex
\section{Review of the junction conditions for thin shells} \label{sec:shellreview}

We now review the the junction conditions \cite{Israel:1966rt} with the shell.  For discussions see \cite{Keranen:2015fqa,Bezrukov:2015ufa,Chandra:2022fwi}. 

Let's consider the Einstein Hilbert action for Euclidean-AdS$_{d+1}$ with a pressure-less shell made from dust of particles
\begin{align}
S &= - \frac{1}{16 \pi G_N} \int_{\mathcal{M}} d^{d+1}x \sqrt{g} \left( R  + d(d-1) \right) + \frac{1}{8\pi G_N} \int_{\partial \mathcal{M} } d^dx \sqrt{h} \left(K - K_0 \right) - \int m(r)  d\ell
\end{align} Here we have fixed the AdS radius to 1.  The metric inside ($-$) and outside ($+$) of the shell are given as
\begin{equation}
ds^2 = f_{\pm} d\tau_\pm^2 + \frac{dr^2}{f_\pm} + r^2 d\Omega^2_{d-1}.  
\end{equation} We can go to Lorentzian by a Wick rotation $\tau_\pm \to  i t_\pm$ .  The metric is continuous across the shell with the constraint 
\begin{equation}
 d\ell^2 = \left( f_+ \dot{\tau}_+^2 + \frac{\dot{r}^2}{f_+} \right) d\ell^2 = \left( f_- \dot{\tau}_+^2 + \frac{\dot{r}^2}{f_-} \right) d\ell^2.
\end{equation}  

The energy density profile is give by $m(r)$.  Following \cite{Bezrukov:2015ufa} it is useful to insert 1 and rewrite the action on the shell as 
\begin{equation}
\int m(r) ~ d\ell = \int d^2 y ~\frac{d \Omega_{d-1}}{\omega_{d-1}} \int ~ d\ell~ m(r) ~\sqrt{ g_{mn} \dot{y}^m \dot{y}^n} ~  \delta^{(2)} (y-y(\ell)) 
\end{equation} where $y^m \in \{\tau, r\}$ and $g_{mn}$ is the metric along those directions.  The volume of the unit sphere is $\omega_{d-1}$.  The component of the stress tensor along the sphere vanish and we obtain
\begin{equation}
T^{mn}_{shell} = \dot{y}^m \dot{y}^n \frac{m(r)}{\omega_{d-1} r^{d-1}} \int d\ell ~\delta^{(2)} (y-y(\ell))   
\end{equation}  
The junction conditions are given in terms of the extrinsic curvatures as
\begin{equation}
\left[K\right]_{ab} - \left[K\right] h_{ab} = - 8\pi G_N T_{ab} 
\end{equation} where $\left[K\right] = K_+ -K_-$. We pick the normal of the shell to be outwardly pointing.  The $\{a,b\}$ indices correspond to coordinates on the shell, and $T_{ab}$ is the pull-back of the bulk stress-tensor. 
The extrinsic curvatures evaluate to 
\begin{equation}
K_{ij} =  \dot{\tau}  ~r~  f(r) \hat{g}_{ij}, \qquad K_{\ell \ell} =   \frac{\partial_\ell (\dot{\tau} f)}{\dot{r}} 
\end{equation}  where $\hat{g}_{ij}$ is the metric on the unit sphere.  All together the equations of motion become
\begin{equation}
\dot{\tau}_- f_- - \dot{\tau}_+ f_+ = \frac{8\pi G_N ~m(r)}{(d-1) \omega_{d-1}~ r^{d-2}}, \qquad \partial_\ell \left( r^{d-2} \left[\dot{\tau}_- f_- - \dot{\tau}_+ f_+\right] \right) =0.  
\end{equation}  The second equation fixes $m(r) = m$ to a constant.  This equation is simply the continuity equation of the stress tensor on the shell.  Indeed the second equation comes from the components of the junction equation along the sphere.  For an ideal fluid this would fix the pressure on the shell, which is vanishing in our setup.  

We solve for the profile of the shell from the relation
\begin{equation}
4 ~ \dot{r}^2 ~ \left(\frac{\tilde{m}}{r^{d-2}}\right)^2 - 4 f_+ f_- + \left(\left(\frac{\tilde{m}}{r^{d-2}}\right)^2 - f_+ -f_- \right)^2 =0, \qquad \tilde{m} \equiv \frac{8\pi G_N }{(d-1) \omega_{d-1}}m.  
\end{equation}  
It is useful to recast the equations as
\begin{equation} \label{eq:rdothd}
 \dot{r}^2 = f_\pm - a_\pm^2, \qquad \dot{\tau}_\pm f_\pm =  a_\pm \equiv \mp \frac{ \left(\frac{\tilde{m}}{r^{d-2}}\right)^2 \pm (f_+ - f_-)}{2\frac{\tilde{m}}{r^{d-2}}} . 
\end{equation}

%% file: sec_sphericalSym.tex
\section{Action of spherically symmetric thin shell solutions} \label{sec:sphericalSym}

 
 
 
 
 In this section we will discuss the computation of the action for general spherical symmetric thin shell solutions. Here we will try to be as general as possible and consider the problem in general two dimensional "entropion" or dilaton gravities.

 \subsection{The action} 
 
 Any problem involving spherically symmetric shells reduces to a two dimensional gravity problem since nothing will depend on the angular directions. Furthermore, any two dimensional gravity action can be written, up to field redefinitions, as ``entropion'' gravity\footnote{This is sometimes called ``dilaton'' gravity, which is a bad name since the field $s$, in two dimensions,  is not associated to dilatations but to the entropy.   } 
 \be \la{ActEG}
 - I_{\rm grav} = \int \sqrt{g } \left[ { s  \over 4 \pi } R + {\cal T }(s) \right] 
 \ee 
 where $s$ is a field and $\mathcal{T}(s)$ a general function of that scalar field. This scalar field is equal to $s = A/4 G_N$ where $A$ is the area of the sphere. 
 The general solutions have an isometry, which we make manifest as shifts of a coordinate $\tau$. In addition we choose a coordinate $\rho$ set by the value of the scalar field,  $s=\rho$. We use $\rho$ rather than $r$ to distinguish it from the radial coordinate in the higher dimensional discussion. The equations then imply that 
 %
 %
 %
 \def\h{\hat f}  
 \be \la{Coord}
 ds^2 =  \h d\tau^2  + { d\rho^2 \over \h } ~,~~~~~~~~~s =\rho ~,~~~~~~~~\h' = 4 \pi {\cal T}(\rho)  
 \ee    
 or 
 \be \label{fE}
 \h = 4 \pi \left[  {\cal E}(\rho) - {\cal E}(\rho_h) \right] ~,~~~~~{\rm with} ~~~~ {\cal E}'(\rho) = {\cal T}(\rho)
 \ee 
 where $\rho_h$ is where $\h=0$ and corresponds to a horizon of the solution. The constant of integration in \nref{Coord} was chosen to ensure this.  We have   written the equations in the Euclidean signature. 
 The thermodynamic quantities are then equal to 
 \be \la{Trel}
 S = s_h = \rho_h ~,~~~~~~~~T = {\cal T}(\rho_h) ~,~~~~~~~~~~E = {\cal E}(\rho_h) 
 \ee 
 so that we see that the function ${\cal T} (s)$ appearing in \nref{ActEG} is simply the temperature as a function of the entropy \cite{Maldacena:2019cbz,Witten:2020ert}.
 
 To this action, we can add the thin shell action 
 \be 
 - I_{\rm shell} = - \int \hat m(s) d\ell 
 \ee 
 where $\ell$ is the proper length and $\hat m(s)$ is the proper mass which could depend on the field $s$. Recall that in higher dimensions $s$ is related to the area of the sphere and the shell action could depend on it. 
 
 We can derive the junction conditions from this 2d point of view by looking at the equations of motion for the total action $I_{\rm grav} + I_{\rm shell}$ in the vicinity of the shell. These are easier to obtain in conformal gauge for the metric $g_{\mu \nu} = \eta_{\mu \nu} e^{ 2 \hat \omega}$. For a shell along $x^0$ in these coordinates the relevant terms near the shell are 
 \be \la{Actc}
 - I \sim  \int dx^0 dx^1 {   s \over 2 \pi } (-\partial^2 \hat \omega)  - \int dx^0 \hat m(s) e^{\hat \omega} 
 \ee 
 giving the equations 
 \be \la{jumc}
 { 1 \over 2 \pi }  \partial_{1} \hat \omega |^+_-  + \partial_s \hat m(s)e^{\hat \omega} =0~,~~~~~~~~~ { 1 \over 2 \pi } \partial_{ 1} s |^+_-  + \hat m(s) e^{\hat \omega} =0
 \ee 
where $\pm$ indicate the two sides of the shell. 
 We can write \nref{jumc} in a coordinate invariant way as
 \be \la{JunCov}
 { 1 \over 2 \pi } ( K_+ - K_-) + \partial_s \hat m(s) =0 ~,~~~~~ {1 \over 2 \pi } \partial_n s |^+_- + \hat m(s) =0 .
 \ee 
 
The first equation in \nref{jumc} implies also a delta function in the curvature localized at the shell. When we integrate the action this leads to an extra term of the form  
\begin{equation} \la{CurvDel}
	\int \sqrt{g}\,\frac{s}{4\pi} R =\int \sqrt{g}\,\frac{s}{4\pi} \left( - 2 \nabla^2 \hat \omega\right) = \int dx^1\, e^{\hat \rho} s\partial_s \hat m(s) =  \int d\ell \, s\partial_s \hat m(s).
\end{equation} 
 where the first integrals are only around a small strip  including the shell. 
 
   
 In addition, we are demanding that the metric and the field $s$ are continuous along the shell. 
 Going back to our coordinates in \nref{Coord} we notice that the coordinate $\rho$ is the same on both sides of the shell, but $t_\pm$ could be different. In addition, the function $\h$ could be different since there can be different integration constants $\rho_\pm = \rho_{h, \pm}$. Then the second junction condition \nref{JunCov}, together with the continuity equation for the metric implies,     
 \be \la{eomshell}
  \dot{\tau}_+ \h_+ -  \dot{\tau}_- \h_- + 2 \pi \hat m =0 ~,~~~~~~~~~ 1 =  \h_+ \dot \tau_+^2 + { \dot \rho^2 \over \h_+} =   \hat{f}_- \dot \tau_-^2 + { \dot \rho^2 \over \h_-}
 \ee 
 The first junction condition in \nref{JunCov} can the be derived from these and the bulk equations. 
  
 From (\ref{eomshell}) we can derive that 
 \begin{equation} \label{rdote}
  \dot \rho^2 = \h_\pm - a_\pm^2, \qquad \dot \tau_\pm \h_\pm = a_\pm =  \mp { (2\pi \hat m)^2 \pm (\h_+ -\h_- ) \over 4\pi \hat m }  \end{equation} This is consistent with the higher dimensional junction conditions in \eqref{eq:rdothd}.
%
  

  \subsection{Evaluating the action} 
  
  We now turn to the problem of evaluating the action. We will be interested in configurations with fixed energy at the boundary, rather than fixed time difference. So, we now discuss the corresponding boundary terms. The gravitational action need to be supplemented by the following boundary terms 
  \be \la{bdyterm}
  - I_{\rm grav \, bdy} = \int_{\rm bdy} { s_b \over 2 \pi } K + \beta E =  \int_{\rm bdy} { s_b \over 2 \pi } K +  E \int d\tau  
  \ee 
 where we are cutting out the boundary at radial cutoff $\rho_b=s_b$.

The total action is given by
\begin{equation}\label{logZtotal}
\begin{aligned}
	\log Z & = - I_{\rm grav} - I_{\rm grav\, bdy} - I_{\rm shell} \\
	& = \int \sqrt{g } \left[ { s  \over 4 \pi } R + {\cal T }(s) \right]  + \int_{\rm bdy} { s_b \over 2 \pi } K +  E \int d\tau  - \int \hat m(s) d\ell  
\end{aligned} 
\end{equation}
The action has a divergence when the cutoff is taken to infinity, so we also need to add a suitable counterterm $I_{\textrm{ct}}$ in order to extract the interesting finite part. We will elaborate the explicit form of $I_{\textrm{ct}}$ later.


Using the equations of motion the on shell action can be simplified as follows. 
The bulk terms
\begin{equation}\label{bulkterm}
	\int \sqrt{g} \left[ \frac{s}{4\pi }R + \mathcal{T} (s)\right] =\frac{1}{4\pi }  \int d\tau \, \int d\rho \, (- (\h'\rho)'+2\h')
\end{equation}
become a total derivative in $\rho$ and can be integrated into boundary terms. There are several different types of "boundaries" we can have. The simplest is if the integral in $\rho$ starts at the horizon. In this case we get contribution
\begin{equation} \la{EntCon}
	{ \Delta \tau \over 4 \pi } (\h'\rho)|_{\rho=\rho_h}= \frac{\beta}{4\pi } (\h'\rho)|_{\rho=\rho_h} = s_h. 
\end{equation}
In other words, we get a contribution that is the entropy whenever a fixed point of $U(1)$ action is in the geometry. Note that for $\Delta \tau \to - \beta$ this contribution changes sign. This is what happens when we have negative Euclidean time evolution. 

The second kind of boundary is the cutoff boundary at large radius $\rho_b$, from which we get
\begin{equation}\la{secb}
	\frac{1}{4\pi } \int d\tau \left[ - \h'(\rho_b)\rho_b + 2\h(\rho_b)\right]
\end{equation}
We note that the first term in (\ref{secb}) exactly cancels the extrinsic curvature term in (\ref{bdyterm}). To cancel the divergence from the second term $\h(\rho_b)$, we need to add a counterterm proportional to the length of the boundary
\begin{equation}
	-I_{\textrm{ct}} = -\frac{1}{\sqrt{4\pi }} 2 \sqrt{ \mathcal{E}(\rho_b) } \int  d\tau\,  \sqrt{ \h(\rho_b)} ,
\end{equation}
where we adjusted the coefficient so that the leading term would cancel the last term in \nref{secb}. 
Combining (\ref{secb}) with the boundary terms in (\ref{bdyterm}) as well as the counterterm $I_{\textrm{ct}}$, we get
\begin{equation}
\begin{aligned}
	&  \frac{1}{4\pi } \int d\tau \left[ - \h'(\rho_b)\rho_b + 2\h(\rho_b)\right]
 + \int_{\rm bdy} { s_b \over 2 \pi } K +  E \int d\tau - 2 \sqrt{\mathcal{E}(\rho_b)}  \int d\tau\,  \sqrt{ \h(\rho_b) \over 4 \pi } =\\
 & =    \int d\tau\, 2\mathcal{E}(\rho_b)  
 + E \int d\tau -   \int d\tau\, 2\sqrt{(\mathcal{E}(\rho_b) - \mathcal{E}(\rho_h))\mathcal{E}(\rho_b)} \\
  & = 0 ,\quad \quad {\rm as } ~~~~~~ \rho_b\rightarrow \infty.
\end{aligned}
\end{equation}
In the last step, we used that $E = \mathcal{E}(\rho_h)$ as in (\ref{Trel}). So the conclusion is that we don't get any net contribution from the spatial infinity!

Back to (\ref{bulkterm}), other than the horizon and spatial infinity, we can also pick up contributions from the inner/outer "boundary" at the thin shell. We have
\begin{equation}\label{bulkshell}
\begin{aligned}
	 \frac{1}{4\pi} \int \left( - (\rho \h') + 2 \hat f  \right)(d\tau_- - d\tau_+) & ={ 1 \over 4 \pi } \int (\rho \h') (d\tau_+ - d\tau_-) + \frac{1}{2\pi} \int d\ell \, (\dot{\tau}_- \h_- - \dot{\tau}_+ \h_+) \\
	& = { 1 \over 4 \pi } \int (\rho \h') (d\tau_+ - d\tau_-) +  \int \hat m(s) d\ell . 
\end{aligned}
\end{equation}
where we used (\ref{eomshell}) to go to the second line. The second piece in (\ref{bulkshell}) cancels the last term in (\ref{logZtotal}).

There is one additional contribution from (\ref{bulkterm}) that is easy to miss. This comes from a delta-function contribution in the Ricci scalar localized at the shell which gives, see \nref{CurvDel},  
\begin{equation}
	 	 -I _{\rm delta~function} = \int_{\textrm{near shell}}  {s \over 4 \pi } R  =\int d\ell \, s\partial_s \hat m(s).
\end{equation}
where the integral is over a small region near the shell that picks up the delta function contribution. 

Finally, after collecting all the pieces, we get 
\be \la{logZ}
\log Z = \sum_i s_i + { 1 \over 4 \pi } \int (\rho \h') (d\tau_+ - d\tau_-) + \int d\ell\, \rho \partial_\rho \hat m
\ee 
where $s_i$ is the contribution from each fixed point of the $U(1)$ action. As we explained around (\ref{smallm}), in the limit that the rest mass is small,  $ E\gg \hat m$, we can ignore the last two terms and only the entropy contribution survives.

In order to go to the notation of appendix \ref{sec:shellreview} we need to remember that 
\be 
\rho = {A \over 4 G_N} = { \omega_{d-1} \over 4 G_N} r^{d-1}
\ee 
 and also the two dimensional  metric here is the $r$ and $t$ part of the metric there but rescaled by a factor of ${ d\rho \over dr }$, namely
  \be 
  \hat f d\tau^2 + { d \rho^2 \over \hat f } = { d \rho \over dr } \left[ f d\tau^2 + { dr^2 \over f } \right]  ~~~~~~~ \to ~~~~~\hat f = { d \rho \over d r } f .
  \ee 
  In particular,  we can use this to rewrite the action \nref{logZ} in terms of the variable of appendix A and the main body of the paper. 
  When we do that we also need to remember that $d\ell_{here} = \sqrt{ d \rho \over dr } d\ell_{there}$. Similarly, $\hat m_{here} = \left( {d \rho \over d r } \right)^{-1/2} m_{there}$. In the main body of the paper the total mass of the shell was $r$ independent. However, due to the extra factor the mass in \nref{logZ} is $\rho $ dependent and leads to a non-trivial contribution. 
  Similarly, the term involving $ \rho \partial_\rho \hat f$ can be  rewritten as $ { 1 \over d-1} r \partial_r ( { d \rho \over d r } f ) $. We can then rewrite the resulting expression using \nref{AdSd} to get the formula \nref{actionAdSd}.

\section{Comment about potential barrier}\label{sec:barrier}

 In our calculation in sec. \ref{sec:oneshell}, the black hole is in thermal equilibrium with its environment, so it is convenient to simply extract the radiation (in this case, the shell) from the thermal environment far away rather than extracting it from the near horizon region.

\begin{figure}[h]
    \begin{center}
   \includegraphics[scale=.1]{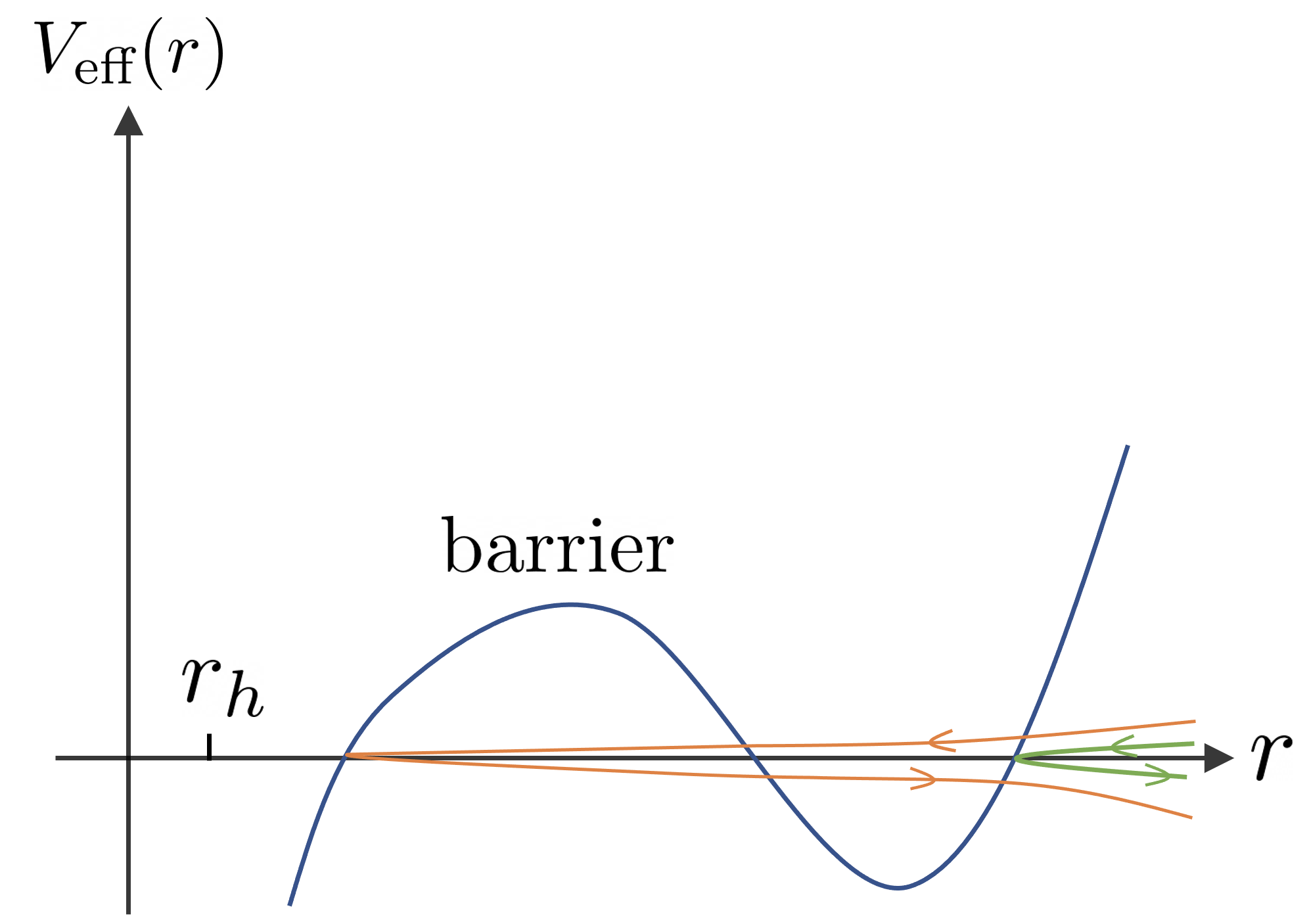}
    \end{center}
    \caption{  We can consider a case where there is a potential barrier for the shell separating the near horizon region ($r\sim r_h$) and far away region. In such cases, we expect it to cost more action if the shell is nucleated near the horizon. }
    \label{fig:potential}
\end{figure}

  To illustrate this point further, we could imagine a situation that there is a potential barrier separating the particle from the near horizon region to infinity. In other words, we can have
  \begin{equation}
  	\dot{r}^2  = V_{\rm eff}(r)
  \end{equation} 
  where the effective potential $V_{\rm eff} (r)$ has multiple zeros outside the horizon, see fig. \ref{fig:potential}. Here $V_{\rm eff}(r)$ is the effective potential for the movement of the shell in the Lorentzian signature. In our Euclidean tunneling solution, the trajectory exists in the classically forbidden region, i.e. where $V_{\rm eff} (r) >0$, see the green curve in fig. \ref{fig:potential}.

   In such cases, we could consider a different solution for the shell where it is nucleated at a location that is closer to the horizon. This is represented by the orange curve in fig. \ref{fig:potential}. Since the shell has to tunnel through an extra potential barrier in order to get to the boundary, we expect that this solution will be further suppressed. 
   
   Of course, in a situation where the black hole is not in thermal equilibrium with its environment, then we have no choice but to extract the particles near the horizon and there will be an extra suppression. 

%% file: sec_JTcheck.tex
\section{Comparison with the exact answer in JT gravity} 
\la{JTCheck}

\subsection{Negative Euclidean evolution in JT gravity}

In the discussion of sec. \ref{sec:TwoShell}, we introduced the quantity $\tau_{AB} (E,E')$, which is the Euclidean time separation between operators $A$ and $B$. In the case of JT gravity, this separation can be worked out explicitly and it takes the following form:
\begin{equation}\la{tauJT}
	\tau_{AB}  = {\beta \over 2 \pi } \, \theta  ~,~~~~~
{\rm with} ~~~~~~\sin { \theta \over2 }  =    \frac{ E' - E + \pi m^2  }{ \sqrt{(E'-E+\pi m^2)^2 + 4\pi E m^2 } }  ~,~~~~~
\end{equation}

When $E'\gg E$, namely the situation considered in fig. \ref{TwoEasy}, we have $\tau_{AB} \approx \beta/2$. On the other hand, as we take $E$ larger, the right hand side of (\ref{tauJT}) becomes negative. Therefore the analytic continuation of  $\tau_{AB} (E,E')$ to the regime $E\gg E'$ becomes negative and eventually approaches $-\beta/2$.

\subsection{Comparing the action}

  Here we evaluate the action (\ref{logZ}) for the specific case of JT gravity, where we can compare it to the exact answer from Schwarzian theory \cite{Mertens:2017mtv}.
  
  We look at two cases, first with $E< E'$, where the geometry doesn't involve negative Euclidean evolution, see fig. \ref{TwoEasy}. The second case involves $E>E'$, which is the case we are interested in and does involve negative Euclidean evolution, see fig. \ref{TwoHard}. We'll see that in both cases the classical action agrees with the exact answer.

There are some common features of the both cases. Since here we are treating JT gravity as a two dimensional theory rather than a dimensional reduction of higher dimensional theory, the mass function $m$ is independent of the radial coordinate. Therefore, the total action simplifies to 
\begin{equation}\label{JTlogZ}
	\log Z = \sum_i s_i + { 1 \over 4 \pi } \int (\rho  \hat{f}') (d\tau_+ - d\tau_-) 
\end{equation} 
where the value of $s_i$ depends on the fixed points of $U(1)$ isometry we have in the geometry. It can come with a minus sign if it is located in the region where we have negative Euclidean evolution.  In JT gravity, we have
\begin{equation}
	\hat{f}_\pm = \rho^2 - \rho_{\pm}^2. 
\end{equation}
In fig. \ref{TwoEasy}, $\rho_+$ corresponds to the entropy of black hole with energy $E$, while $\rho_-$ corresponds to the one with $E'$. In fig. \ref{TwoHard}, it is reversed.
Applying (\ref{rdote}) to the specific case of JT gravity, we have
\begin{equation}
\begin{aligned}
& 	\dot{\rho}^2 = \rho^2 - \rho_{t}^2, \quad  \hat{f}_\pm \dot{\tau}_\pm = a_\pm = \frac{\rho_+^2 - \rho_-^2 \mp \tilde{m}^2}{2\tilde{m}},\\
& \rho_t^2 = a_\pm^2 + \rho_\pm^2 = \frac{ (\tilde{m}\pm i\rho_+ \pm i\rho_-)}{4\tilde{m}^2}, 
\end{aligned}
\end{equation}
where we've defined $\tilde{m}\equiv 2\pi m$. Note that $(\tilde{m} \pm i\rho_+ \pm i\rho_-)$ means the product over four terms with different choices of signs.  

We can evaluate the second term of (\ref{JTlogZ}) as follows
\begin{equation}\label{pluscon}
\begin{aligned}
	\frac{1}{4\pi} \int  \rho \hat{f}'_+ d\tau_+ & = \frac{1}{4\pi} \int \rho 2\rho   \frac{\dot{\tau}_+}{\dot{\rho} }d\rho  = \frac{2}{4\pi} \int_{\rho_{t}}^{\rho_c} d\rho\, \frac{2 a_+ \rho^2}{ (\rho^2 - \rho_+^2)\sqrt{\rho^2 - \rho_t^2 } } \\
	& = \frac{1}{2\pi} \left[\int_{\rho_{t}}^{\rho_c} d\rho\, \frac{2 a_+ }{ \sqrt{\rho^2 - \rho_t^2 } } + \int_{\rho_{t}}^{\rho_c} d\rho\, \frac{2 a_+ \rho_+^2}{ (\rho^2 - \rho_+^2)\sqrt{\rho^2 - \rho_t^2 } } \right] \\
	& = \frac{1}{2\pi} \left[ 2a_+ \left(\log \rho_c - \log \rho_t + \log 2 \right)  +  i\rho_+ \log \left( \frac{a_+ - i\rho_+}{a_+ + i\rho_+}\right) + \mathcal{O}\left(\frac{1}{\rho_c}\right)\right]
\end{aligned}
\end{equation}
Replacing $+$ by $-$ we get the contribution from the other term in (\ref{JTlogZ}). The terms that diverges as $\log \rho_c$ can be canceled out by a counterterm 
\begin{equation}
	I_{\textrm{ct}} = - \frac{1}{\pi} (a_+ - a_-) \log \rho_c = \frac{1}{\pi }m \log \rho_c.
\end{equation}
Different ways of regulating might lead to answers that differ by linear terms in $m$, so we won't keep track of such terms in below. Therefore, plugging (\ref{pluscon}) back into (\ref{JTlogZ}), we get
\begin{equation}\label{JTfin}
\begin{aligned}
	\log Z & = \sum_i s_i  + 2\times  \frac{1}{2\pi} \left[   2 (a_- - a_+)\log \rho_t  +  i\rho_+ \log \left( \frac{a_+ - i\rho_+}{a_+ + i\rho_+}\right) - i\rho_- \log \left( \frac{a_- - i\rho_-}{a_- + i\rho_-}\right) \right] \\
	& = \sum_i s_i + \frac{1}{\pi} \left\{ \tilde{m} \log \frac{(\tilde{m} \pm is_+\pm s_-)}{\tilde{m}^2} + is_+ \log \left[\frac{ (\tilde{m} +is_+)^2 + s_-^2}{(\tilde{m} - is_+)^2 + s_-^2}\right] - is_- \log \left[\frac{ (\tilde{m} - is_-)^2 + s_+^2}{(\tilde{m} + is_-)^2 + s_+^2}\right] \right\}.
\end{aligned}
\end{equation}
where we have a factor of two in the second term since we have two particle trajectories. We also used that $\rho_\pm = s_\pm$.

We now specify to the two situations in fig. \ref{TwoEasy} and \ref{TwoHard} and compare with the exact answer.

\subsubsection{$E<E'$}

In this case, the exact answer from Schwarzian theory gives 
\begin{equation}\label{exact}
	\log Z_{\rm exact} = \log \left[ \rho(s')^2 \rho(s) \frac{\Gamma\left(m \pm i \frac{s}{2\pi } \pm i\frac{s'}{2\pi }\right)^2}{ \Gamma(2m)^2}\right], \quad \rho(s)\propto s\sinh s 
\end{equation}
We have $E = s^2/(4\pi), E' = s'^2/(4\pi )$.\footnote{Our convention differs from \cite{Mertens:2017mtv}. We have $E_{\textrm{here} } =\pi E_{\textrm{there}}$, $s_{\textrm{here}} = 2\pi k_{\textrm{there}}$ and our $m$ is $\ell$ there.} In order to compare with the classical answer, we expand it in the semiclassical regime $s'\gg s\gg \tilde{m} \gg 1$, 
\begin{equation}
\begin{aligned}
	\log Z_{\rm exact} & \approx 2 s' +  s + \frac{1}{\pi} ( \tilde{m} \pm is \pm is' )  \log ( \tilde{m} \pm is \pm is' ) -\frac{2}{\pi} \tilde{m} \log (2\tilde{m}) \\
	& \approx  s + \frac{1}{\pi }\tilde{m} \log  ( \tilde{m} \pm is \pm is' )  - \frac{2}{\pi} \tilde{m} \log \tilde{m} + \frac{is}{\pi } \log \left[\frac{ (\tilde{m} +is)^2 + s'^2}{(\tilde{m} - is)^2 + s'^2}\right] -\frac{i s'}{\pi }\log \left[\frac{ (\tilde{m} - is')^2 + s^2}{(\tilde{m} + is')^2 + s^2}\right]
\end{aligned}
\end{equation}
We see that it agrees with what we had in (\ref{JTfin}), after we identify $s_+ = s',s_- = s$. Here we only have one $U(1)$ fixed point in the geometry, associated with the black hole with energy $E$, so we have $\sum_i s_i = s$.

\subsubsection{$E>E'$}

In this case, we expand (\ref{exact}) in a different regime $s\gg s'\gg \tilde{m}\gg 1$. We have
\begin{equation}
\begin{aligned}
	\log Z_{\rm exact} & \approx 2 s' +  s + \frac{1}{\pi} ( \tilde{m} \pm is \pm is' )  \log ( \tilde{m} \pm is \pm is' ) -\frac{2}{\pi}\tilde{m} \log (2\tilde{m}) \\
	& \approx  2s' -s + \frac{1}{\pi }\tilde{m} \log  ( \tilde{m} \pm is \pm is' )  - \frac{2}{\pi} \tilde{m}\log \tilde{m} \\
	& \quad + \frac{is}{\pi } \log \left[\frac{ (\tilde{m} +is)^2 + s'^2}{(\tilde{m} - is)^2 + s'^2}\right] -\frac{i s'}{\pi }\log \left[\frac{ (\tilde{m} - is')^2 + s^2}{(\tilde{m} + is')^2 + s^2}\right]
\end{aligned}	
\end{equation}
It agrees with (\ref{JTfin}) after identifying $s_+ = s,s_- = s'$. As opposed to the previous case, we have three fixed points in the geometry, two coming from the $E'$ black hole and one coming from the $E$ black hole. The fixed point of the black hole with energy $E$ is located in a region with negative Euclidean evolution, so it contributes with a minus sign. Therefore, $\sum_{i}s_i = 2s' - s$, agreeing with the exact answer.

Note that for the topological term we get simply $e^{S_0}$ for both diagrams in figure \ref{TwoEasy} and \ref{TwoHard}. 
We would also get the same topological factor if we wrote the final answer for the diagram of figure \ref{TwoHard} as 
$ \log Z \sim  2 (S_0 + s') - (S_0 + s) $.


%% file: sec_QNM.tex
   \section{Bow tie, boost evolution and quasinormal modes} 
  \la{QNM}


In our construction of the wormhole geometry we see that the middle of the geometry is given by a "bow tie", see figure \ref{TwoShellsBetter}. Even though our whole geometry is smooth, we could wonder whether it is possible to single out the bow tie geometry, fixing boundary conditions along the red and blue dashed lines in fig. \ref{TwoShellsBetter} and define matter partition function on it. In other words, we want a prescription for computing the transition amplitude $ _b\langle \Psi_{B B^\dagger } | e^{ - i t H } | \Psi_{A A^\dagger } \rangle_b $ in (\ref{amp}) on the bow tie.

Naively the geometry seems singular since the bifurcation surface is a fixed point under the boost evolution. 
However, a way to avoid the singularity is to do an analytic continuation as in \cite{Saad:2018bqo}. Writing the metric near the horizon as   
\be 
 ds^2 = - r^2 dt^2 + dr^2 
 \ee 
 with the right side with $r>0$ and the left side with $r<0$, the analytic continuation prescription is
 \be \label{prescription}
  r \to  r - i \epsilon 
  \ee 
Another way to say it is that we do not go through the singularity $r=0$ directly, but instead go through the lower half plane of the complex $r$ plane.\footnote{In \cite{Saad:2018bqo} the prescription is to go through the upper half plane. We believe that the difference comes from that \cite{Saad:2018bqo} considers the boost evolution as $e^{iHt}$, while we are define it in the ordinary way $e^{-iHt}$. }
  
  
  This analytic continuation has the property that the boost evolution leads to a Hamiltonian whose eigenvalues have negative imaginary parts.   In fact the eigenstates relevant to this evolution are the same as the quasinormal modes, as we explain in more detail in appendix \ref{BoostQuasi}. In other words, the eigenfrequencies obey 
  \be 
   \omega = ({\rm real } ) - i ( {\rm positive } ) \la{Egf}
   \ee 
   So that the evolution of the perturbations under the time shift $t$ will be suppressed as 
   \be 
    \exp\left[ - ( {\rm positive} ) t \right] .
    \ee 
    

\subsection{Boost evolution and quasinormal modes}\label{BoostQuasi}
 
 
 In this section we would like to argue that the eigenvalues of the boost evolution, using the prescription in \nref{prescription}, are given by the quasinormal mode frequencies of the black hole. 
 
 We will argue this by noting that the problem that computes the quasinormal modes is mathematically identical to the problem of computing the eigenvalues of the boost operator, subject to the prescription (\ref{prescription}). 
 
 We will consider a general black hole in $AdS$. The quasinormal modes are solutions of the bulk wave equation in the region outside the horizon that have definite frequency under   the Schwarzschild time. For simplicity, we normalize the time direction so that $\beta = 2\pi$, since what we are about to explain does not depend on the normalization of time. 
 \be 
 \la{WfAn} 
 \phi = e^{ - i \omega t } F_w(r)
 \ee 
  The modes obey a Dirichlet boundary condition at infinity. This selects a single solution of the wave equation, up to an overall normalization. Near the horizon they also  obey an ingoing boundary condition. In order to specify  more clearly this boundary condition, we expand the metric near the horizon as 
 \be 
  ds^2 = dr^2 - r^2 dt^2 + \cdots = - dX^+ dX^- + \cdots ~,~~~~~~X^\pm = \pm r e^{\pm t } 
  \ee  
  The solutions of the wave equation near $\rho=0$ go as $\rho^{\pm i   \omega } $. It is useful to rewrite the solutions near the horizon in terms of $X^\pm$ as 
  \be 
  e^{ - i   \omega t } r^{ i \omega } \sim (X^-)^{ i  \omega } ~,~~~~~~~
  e^{ - i   \omega t } r^{ -i \omega } \sim (X^+)^{- i  \omega }
 \ee 
 Notice that the first solution is singular in the future horizon where $X^-=0$ and the second  solution is singular in the past horizon where $X^+=0$. The quasinormal mode condition selects the solution that are non-singular in the future horizon, therefore it selects the second kind of solutions. 
  The function of $r$ that obeys the Dirichlet boundary in \nref{WfAn} behaves as 
  \be \la{SmrBh}
  F\sim  A(\omega) r^{ i   \omega } + B(\omega) r^{ - i   \omega } 
  ~,~~~~ r \sim 0 
  \ee 
  The quasinormal mode condition is 
  \be 
  A(\omega ) =0
  \ee 
  which gives a discrete set of frequencies. These frequencies can be complex because the boundary conditions are not invariant under complex conjugation. 
 
 We now consider the problem of determining the eigenvalues under $H$ evolution. More precisely, we imagine that we have some sources to the past, and we want to expand them as we go to the future, $t\gg 0$. For this problem we now have two sides. The solution obeying the Dirichlet boundary conditions can be written as 
 \be 
  \phi_r = e^{ - i \omega t } F_\omega(r) ~,~~~{\rm for } ~~ r> 0 ~;,~~~~~~~\phi_l = \Lambda e^{ - i \omega t } F_\omega (|r|) ~,~~~{\rm for } ~~r <0
 \ee 
 where $F_\omega$ is the solution of the Dirichlet problem on one side (unique up to an overall scale). We take the same function on both sides, and we allowed an arbitrary relative normalization $\Lambda$ between the two sides. 
 
 We now demand that the function obeys the analytic continuation property (\ref{prescription}). This means that we want that, for $r>0$, 
 \be \label{quasidev}
  \phi_r( e^{ -i \pi } r) = \phi_l(-r) ~~~\longrightarrow ~~~~   A(\omega) e^{ \pi \omega } r^{ i   \omega } + B(\omega) e^{ -\pi \omega } r^{ - i   \omega }  =   \Lambda \left[  A(\omega) r^{ i   \omega } + B(\omega) r^{ - i   \omega }\right] 
  \ee 
  Equating the corresponding powers we find 
  \be \la{CondiQN}
  A e^{\pi \omega} = \Lambda A ~,~~~~~~~~~~B e^{ -\pi \omega } = \Lambda B 
  \ee 
  For generic $\omega$ we cannot obey both equations. So, one of the equations must be trivial. We should have one of two conditions 
  \be \la{AandBcond}
  A(\omega)=0 ~,~~~~~~~~{\rm or }~~~~~~~~~~B(\omega )=0
  \ee 
  But these are precisely the conditions we have obtained above for the quasinormal modes. To be precise, $A(\omega)=0$ gives us the solution that are regular in the future horizon, while $B(\omega)=0$ gives us the solutions that are regular in the past horizon. 
 
  We could also imagine obeying \nref{CondiQN} by taking $\omega =  -i n $, with integer $n$ and $\Lambda = e^{i\pi n}$. In this case the  two exponentials in $r$  differ by an even  integer.  This means that the solutions will not quite have the form \nref{SmrBh} generically.  Instead there will be a solution that starts with $r^{|n|}$ plus positive powers and another that starts with $r^{-|n|}$ and contains an $r^{|n|} \log r $ term in its power series expansion around $r=0$.  This means that we need to set to zero the coefficient of this second solution, which generically (for general black holes) will not happen since we are already fixing $\omega$. Note that for special black holes we can have quasinormal modes with $\omega = i n$, such as the ones in $AdS_2$ or $AdS_3$. 
  
  When we are looking at the eigenvalues of $H$, we want to put sources in the past horizon, so that solutions might be singular there, but are regular in the future horizon. So we select the same condition as the quasinormal mode one.

%% file: sec_distanceshell.tex
\section{Distance between the shells at large $t$}\la{Distance}
 
In this appendix we discuss the distance between the $AA^\dagger$ shell and the $BB^\dagger$ shell in the geometry in fig. \ref{TwoShellsBetter}. We will first separate operators $A$ and $B$ (and similarly for operators $A^\dagger$ and $B^\dagger$) by an almost Lorentzian time $\tau + it,\, t\gg \beta$ with $\tau > 0$, and eventually continuing $\tau$ to $\sim - \beta(E)/2$.

\begin{figure}[h]
    \begin{center}
   \includegraphics[scale=.25]{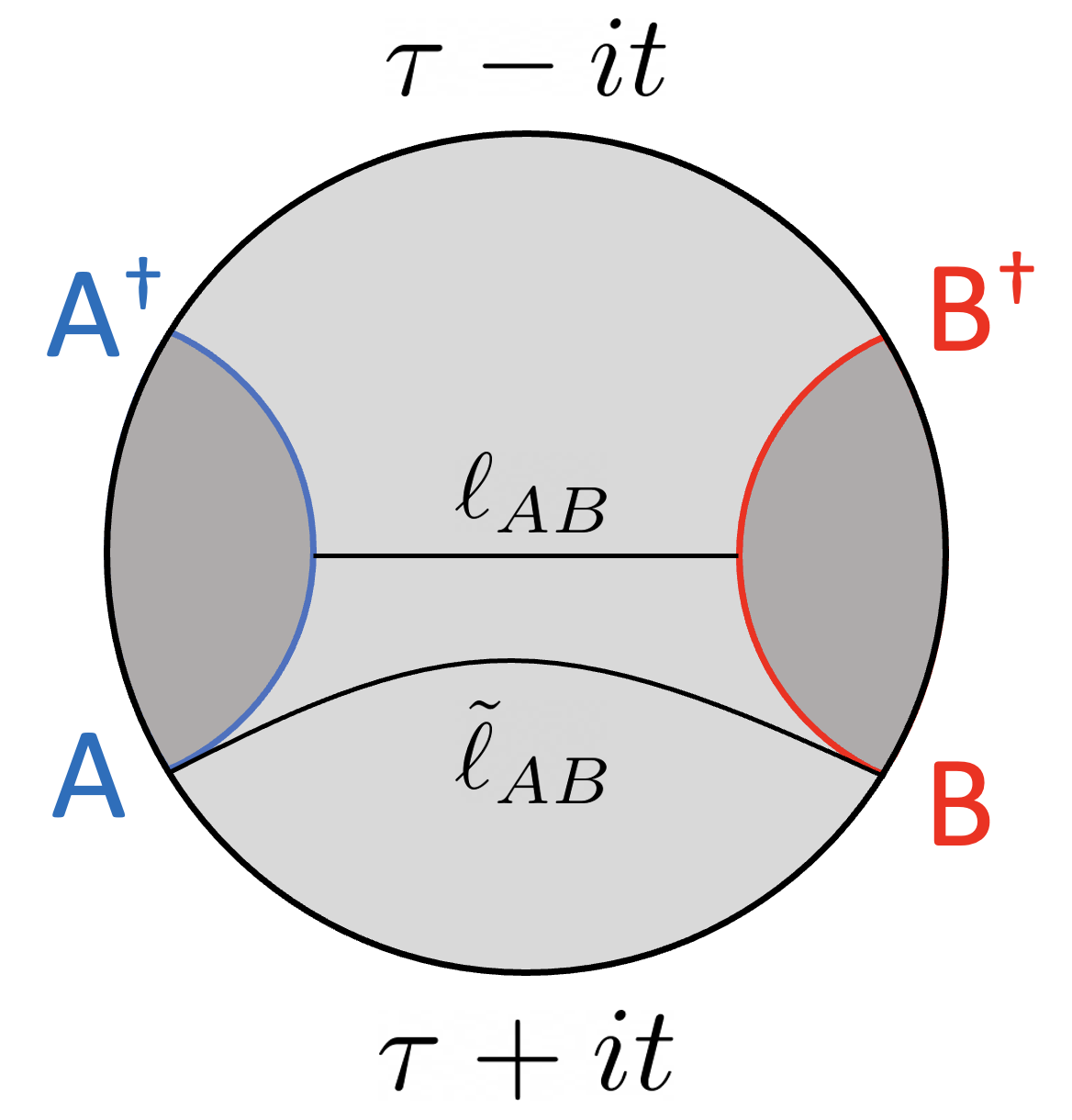}
    \end{center}
    \caption{Computing the distance between the two shells.}
    \label{fig:Distance}
\end{figure}

The configuration is shown in fig. \ref{fig:Distance}. In principle, we would like to compute the shortest geodesic distance between the two shells, denoted by $\ell_{AB}$ in the diagram. The computation of it is a bit more complicated since the end points of the geodesic are at finite radius. However, we can instead compute $\tilde{\ell}_{AB}$, which is the geodesic distance between the operators $A$ and $B$. In higher dimensions, we fix the two end points of the geodesic to be at the same point on the sphere. $\tilde{\ell}_{AB}$ is a good proxy to $\ell_{AB}$ since in our final geometry, the shells are very close to the boundary of the spacetime. The main difference in $\tilde{\ell}_{AB}$ and $\ell_{AB}$ simply comes from the asymptotic regions, which gives rise to a divergence in $\tilde{\ell}_{AB}$ that we shall subtract. After the subtraction, we expect $\tilde{\ell}_{AB}$ and $\ell_{AB}$ to have similar dependence on $t$.

We will do the calculation explicitly for geodesics with zero angular momentum in AdS$_5$, though we expect that the feature that the distance remains large even with negative Euclidean evolution is general.\footnote{One can also verify the same claims in the case of JT gravity, using for example formula (A.15) in \cite{Hsin:2020mfa}.} The distance we are interested in here has been studied in the context of two point functions in eternal black holes. We will follow the discussion in \cite{Fidkowski:2003nf} where the large $t$ limit of the distance was analyzed. The geodesic is parametrized by a single parameter $\xi$, which is determined by the time separation $\tau + it$ by 
\begin{equation}\label{timexi}
	\tau + it = \int dr  \frac{\xi}{\sqrt{ f(r) - \xi^2}f(r)}, \quad f(r) = 1+ r^2 - \frac{r_h^{2} (1+r_h^2)}{r^{2}}.
\end{equation}
The AdS radius is set to one in this calculation. After determining the parameter $\xi$, the distance is computed via
\begin{equation}
	\tilde{\ell}_{AB} = \int \frac{dr}{\sqrt{f(r) - \xi^2}} . 
\end{equation}
Reference \cite{Fidkowski:2003nf} worked out the explicit relation between $\tau+it$ and $\xi$ 
\begin{equation}\label{texplicit}
\begin{aligned}
	\tau+it  &  =   \frac{\beta}{2} - \frac{i}{(2r_h^2+1)}  \left[ r_h \log \left( \frac{(1-\xi^2) - 2i\xi r_h + 2 r_h^2}{ \sqrt{ (1-\xi^2)^2 + 4 r_h^2 + 4 r_h^4 }}\right) \right.\\
	&  \quad\quad\quad\quad\quad\left. - i \sqrt{r_h^2+1} \log \left( \frac{(1+ \xi^2) -2 \xi \sqrt{r_h^2 + 1} + 2r_h^2 }{\sqrt{ (1-\xi^2)^2 + 4 r_h^2 + 4 r_h^4 }  } \right)\right]
\end{aligned}
\end{equation}
Given a value of $\tau + it$, this equation does not determine $\xi$ uniquely. \cite{Fidkowski:2003nf} studied the specific case $\tau = \beta/2$ and found that complex solutions of $\xi$ gives the dominant calculation. These correspond to complex geodesics that do not live in the real sections of the spacetime. In fact, similar conclusion holds for $\tau < \beta/2$. When $t$ becomes large, $\xi$ approaches a complex value 
\begin{equation}\label{xistar}
	\xi_* = \sqrt{1 - 2i r_h \sqrt{1+r_h^2}}, \end{equation}
which is a solution of the equation
\begin{equation}
 (1-\xi^2)^2 + 4 r_h^2 (1+r_h^2)=0.
\end{equation}
The final answer involves summing over this solution and another solution corresponds to $-\bar{\xi}_*$. They will have similar properties, so we focus on one for now.
We can expand (\ref{texplicit}) around $\xi=\xi_*$ and find
\begin{equation}
\begin{aligned}
	t \approx -i\tau +   t_0 - \frac{1}{2(r_h + i\sqrt{r_h^2 + 1})}  \log (\xi - \xi_*)
\end{aligned}
\end{equation}
where $t_0$ is some complex constant, and we get
\begin{equation}\label{xilargeT}
	\xi =\xi_* + e^{- 2\left(r_h + i \sqrt{r_h^2 + 1}\right) (t - t_0 - i\tau)}, \quad t \gg \beta.
\end{equation}
From (\ref{xilargeT}) it becomes clear that by continuing $\tau$ to $\sim -\beta/2$ wouldn't affect the result too much. It will push $\xi$ slightly further away from $\xi_*$, but they are still very close since it is dominated by the dependence on $t$.

The distance $\tilde{\ell}_{AB}$ is then determined by
\begin{equation}
	\tilde{\ell}_{AB}= \int \frac{dr}{\sqrt{f(r) - \xi^2}}  = 2 \log (2r_c ) - \frac{1}{2} \log \left[ (1-\xi^2)^2 + 4 r_h^2 (1+r_h^2)\right] 
\end{equation}
where we dropped terms that vanish as we take the cutoff $r_c$ to infinity. The term $2\log (2r_c)$ is just the usual UV divergence, which  is canceled by counterterms. The second term tells us how $\tilde{\ell}_{AB}$ depends on $t$. By expanding it around $\xi=\xi_*$, we find
\begin{equation}
	\tilde{\ell}_{AB} \sim -\frac{1}{2} \log (\xi- \xi_*) \sim  \left( r_h + i \sqrt{r_h^2 + 1}\right) (t-t_0 - i\tau), \quad t\gg \beta,
\end{equation}
so we see that taking $\tau$ from positive to $\sim -\beta/2$ slightly shortens the distance, but the distance remains very large since it is dominated by the term linear in $t$. 
Note that the distance has an imaginary part that grows with $t$. This is related to the oscillatory behavior of the quasinormal modes in the eternal black hole background \cite{Fidkowski:2003nf,Festuccia:2005pi}.